 \documentclass[iop,apj,appendixfloats,revtex4]{emulateapj}
 
\usepackage[backref,breaklinks,colorlinks,citecolor=blue]{hyperref}
\usepackage[all]{hypcap} %Links go to figures;breaks on deluxetables       

\usepackage{color}

\newcommand{\hst}{{\it HST}}
\newcommand{\Nfid}{N_\mathrm{fiducial}}
\newcommand{\dls}{D_\mathrm{ls}}
\newcommand{\ds}{D_\mathrm{s}}
\newcommand{\dl}{D_\mathrm{l}}
\newcommand{\zs}{z_\mathrm{s}}
\newcommand{\zl}{z_\mathrm{l}}
\newcommand{\LCDM}{$\Lambda$CDM}

\shorttitle{Lensing systematics}
\slugcomment{ApJ in prep: draft date \today}
\shortauthors{Johnson et~al.}

\begin{document}

%==================================================================================
%   TITLE, AUTHORS, AFFILIATIONS
%==================================================================================
\title{The systematics of strong lens modeling quantified: the effects of constraint selection and redshift information on magnification, mass, and multiple image predictability}

\author{Traci L. Johnson and Keren Sharon}
\affil{University of Michigan, Department of Astronomy, 1085 South University Avenue, Ann Arbor, MI 48109-1107, USA}

\email{tljohn@umich.edu}

%==================================================================================
%   ABSTRACT
%==================================================================================
\begin{abstract}
Until now, systematic errors in strong gravitational lens modeling have been acknowledged but have never been fully quantified. Here, we launch an investigation into  the systematics induced by constraint selection. We model the simulated cluster Ares 362 times using random selections of image systems with and without spectroscopic redshifts and quantify the systematics using several diagnostics: image predictability, accuracy of model-predicted redshifts, enclosed mass, and magnification. We find that for models with $>15$ image systems, the image plane rms does not decrease significantly when more systems are added; however, the rms values quoted in the literature may be misleading as to the ability of a model to predict new multiple images. The mass is well constrained near the Einstein radius in all cases, and systematic error drops to $<2\%$ for models using $>10$ image systems. Magnification errors are smallest along the straight portions of the critical curve, and the value of the magnification is systematically lower near curved portions. For $>15$ systems, the systematic error on magnification is $\sim2\%$. We report no trend in magnification error with fraction of spectroscopic image systems when selecting constraints at random; however, when using the same selection of constraints, increasing this fraction up to $\sim0.5$ will increase model accuracy. The results suggest that the selection of constraints, rather than quantity alone, determines the accuracy of the magnification. We note that spectroscopic follow-up of at least a few image systems is crucial because as models without any spectroscopic redshifts are inaccurate across all of our diagnostics.
\end{abstract}


%==================================================================================
%   INTRODUCTION
%==================================================================================
\section{Introduction}
\label{sec:introduction}
Since the discovery of the first gravitationally lensed arc in the field of cluster Abell 370 nearly three decades ago \citep{Soucail:1988kx}, astronomers have been taking advantage of the lensing magnification boost from massive galaxy clusters to observe the distant Universe. Gravitational lensing has the advantage of achromaticity, making spectral observations of these lensed objects comparable to unlensed sources in the field. Additionally, the strong lensing evidence traces the total mass distribution of the cluster, allowing for accurate reconstructions of the projected mass density of clusters, especially on small scales ($<100$ kpc), where other mass tracing methods are sensitive, yet lack the resolution of strong lensing. Lensing is sensitive only to mass and not to gas physics that can contribute to the uncertainties of mass scaling relations where the observable depends on the state of the hot intercluster medium.

However, one of the largest challenges in exploiting gravitational lensing remains in calibrating these ``cosmic telescopes." Both strong and weak lensing methods have been used extensively to measure the mass distribution of galaxy clusters. Numerous weak lensing surveys have allowed for a deep investigation into the statistical and systematic errors of weak lensing methods \citep{Massey:2013xy,Applegate:2014jk,Shirasaki:2014qf}. Similar analyses for strong lensing have lagged behind those of weak lensing for two main reasons: (1) accurate strong lens models require an exquisite image quality to robustly identify multiple images \citep{Kneib:1996kb}, and (2) the occurrence of strong lensing events is lower than weak lensing, making it more difficult to find strong lensing clusters to model. The {\it Hubble Space Telescope} (\hst) has been the primary workhorse for strong lensing observations since the installation of WFPC2. The probability of finding strong lenses is indeed small \citep{Bartelmann:1998fr,Wambsganss:2004vn}; but as predicted, several hundred strong lensing clusters have been found directly in several optical surveys \citep{Gladders:2003zr,Hennawi:2008mz} and after optical follow-up of clusters found in X-ray-selected \citep{Postman:2012lr} and Sunaev-Zel'dovich effect-selected clusters \citep{Menanteau:2010fu,Bleem:2015gf}. Strong lensing mass estimates of the cores of these clusters, when combined with other proxies for mass at larger scales, will allow for measurements of the mass-concentration relation of galaxy clusters \citep{Gralla:2011kx,Oguri:2012bs,Merten:2015rz} across a range of cluster masses and redshifts. These strong lensing clusters highly magnify numerous galaxies from the peak of cosmic star formation density around $z=2$ \citep{Bayliss:2011gf}, allowing for zoomed-in studies of star formation at a time when the Universe produced most of its stars \citep{Madau:2014qd}. Currently, strong lensing clusters are our best chance of finding the most distant galaxies at $z>8$, which may be responsible for re-ionizing the universe \citep[][to name a few]{Zitrin:2014uq,Atek:2015qv,Coe:2015qf,McLeod:2015nr}. With the ever increasing number of known strong lenses, it will be important to fully understand how well we can quantify both the statistical and systematic errors in modeling strong lensing galaxy clusters.

Estimating the statistical errors in strong lens modeling has become nearly routine with the advancement in lens modeling codes to utilize Markov Chain Monte Carlo (MCMC) algorithms to adequately explore parameter spaces. The literature only mentions instances where systematic errors have been revealed between different models of the same cluster or when comparing new models of a cluster to earlier and obsolete models. For example, \citet{Smith:2009lr} found that including redshift information of the strong lensing constraints has a significant effect on constraining the slope of the mass distribution. Similarly, \citet{Johnson:2014tg} found that the magnification can vary beyond the statistical errors when the redshift information is added for a single system. \citet{Jauzac:2015xy} report an overall increase in magnification values for their new model of Abell 2744 using full-depth Hubble Frontier Field (HFF) data; this effect has many possible causes: adding dozens of new image systems as constraints, including new spectroscopic redshifts, and/or correcting a misidentified image system.

\begin{figure*}
\center
\includegraphics[width=\textwidth]{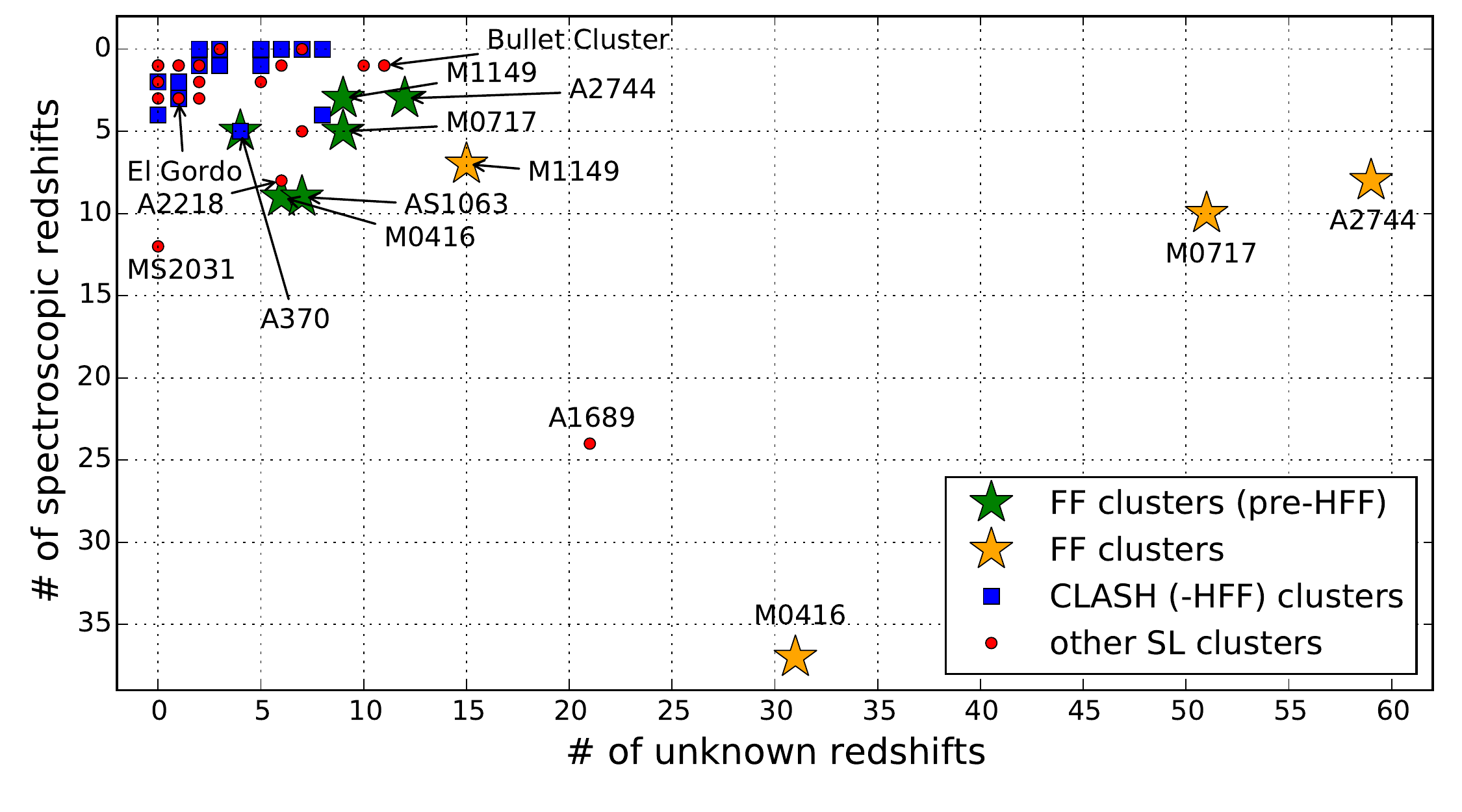}
\caption{Distribution of cluster strong lensing models from the literature, separated into the number of strong lensing image systems with spectroscopic redshifts and those with unknown redshifts. The green stars represent the status of the HFF cluster lens models prior to HFF observations \citep{Johnson:2014tg,Richard:2014gf}, while the yellow stars show the most complete lens models to date on clusters with full HFF data \citep{Jauzac:2014qd,Jauzac:2015xy,Jauzac:2016dn,Caminha:2016fk,Kawamata:2016nr,Limousin:2016ty,Treu:2016lr,}. We also include several other clusters from the literature, including those from the CLASH survey that do not overlap with the HFF clusters \citep{Zitrin:2015lq}. Other clusters include well-known lensing clusters such as Abell 1689 \citep{Diego:2015tg}, the Bullet Cluster \citep{Bradac:2009qd}, El Gordo \citep{Zitrin:2013hl}, Abell 1703 \citep{Limousin:2008lr}, Abell 2218 \citep{Eliasdottir:2007ve}, and many others \citep{Zitrin:2011qy,Sharon:2012ly,Sharon:2014fj,Sharon:2015xe,Bayliss:2014lr,Richard:2007rr,Richard:2010zp,Richard:2015lr}.}
\label{fig:slclusters}
\end{figure*}

While it is true that each individual method has its own systematic errors, all modeling methods are subject to errors due to the availability of constraints. The HFF clusters and Abell 1689, with a wealth of deep multiwavelength imaging and spectroscopy, have unprecedented numbers of image systems identified and have some of the most precise lens models of clusters in existence \citep{Jauzac:2014qd,Jauzac:2015xy,Jauzac:2016dn,Diego:2015tg,Kawamata:2016nr,Treu:2016lr}. Yet, these clusters are seven unique lensing sight lines; in fact, most clusters only have a handful of multiple images (see \autoref{fig:slclusters}). This reality stems from a few factors: (1) the aforementioned clusters are some of the most massive clusters, many showing signs of ongoing growth through mergers \citep{Merten:2011fk,Jauzac:2015qf}, resulting in larger lensing cross-sections, (2) they have some of the deepest \hst\ data allowing for identification of fainter multiple image systems, and (3) they have extensive spectroscopic campaigns that allow for the redshift confirmation of multiple image systems. Determining to what degree systematic errors are induced upon a lens model due to the availability of constraints is a high priority, especially for lower mass clusters, which tend to lens fewer multiple image systems or massive clusters with shallow observations.

In this paper, we begin to address several questions surrounding the topic of systematic errors in lens modeling. How does changing the number of constraining multiple image systems in a lens model affect the accuracy of strong lens models? Similarly, how does increasing the number of spectroscopic redshifts influence a model's accuracy? These questions are timely in this new era of strong lensing where high-quality data are allowing for the most precise (i.e., low statistical uncertainty) models with high numbers of identified lensing constraints. Answering these questions will help guide the lensing community's focus to improve the quality of future strong lensing models. Spectroscopic campaigns are expensive: lensed galaxies, while magnified, are still faint and require long integrations on large telescopes, thus, we must determine their necessity for strong lens modeling. We show that it is critical for strong lens models to have at least a few spectroscopically confirmed redshifts of image systems to dramatically reduce systematic errors.

This paper is organized as follows. We begin with a description of the experimental design in \S\ref{sec:experimental_design}. In \S\ref{sec:slmodeling} we discuss the common methods used in strong lens modeling. We describe the fiducial lens model in \S\ref{sec:lens_model} used for our analysis. In \S\ref{sec:methods} we describe our methodology for quantifying lens modeling systematics and report the results in \S\ref{sec:results}. Finally, we summarize our work in \S\ref{sec:discussion} and discuss plans for future work in \S\ref{sec:future}. We assume a \LCDM\ universe with $\Omega_M=0.3$, $\Omega_\Lambda=0.7$, and $H_0=70\ \mathrm{km\ s^{-1}\ Mpc^{-1}}$. This cosmology yields an angular-physical scale of $1\arcsec=6.104$ kpc to the Ares cluster redshift $z=0.5$.

%==================================================================================
%  EXPERIMENTAL DESIGN
%==================================================================================
\section{Experimental Design}
\label{sec:experimental_design}
The goal of this study is to investigate quantitatively how the selection of strong lensing constraints, i.e., the redshifts and multiple images of strongly lensed galaxies, affect lens models. We do this by generating 350 test models of the same gravitational lens, each test model uses a different random subset of the available constraints. The results of the test models are compared to a fiducial model that uses the full set of constraints as input. In this section, we briefly describe our choices, while a thorough discussion is given in the following sections. 

The best-case-scenario fiducial model is a lens model of the simulated cluster Ares, a model that was initially computed for the purpose of the lens modeling comparison challenge \citep{Meneghetti:2016xe}. The fiducial model is constrained by 66 lensed galaxies with known redshifts. All test models will be compared to this fiducial model, which uses all lensing constraints and the true redshifts of the sources.

The parameter space that is covered by the test models is the number of lensed galaxies with spectroscopic redshifts, and the number of lensed galaxies without spectroscopic redshifts that are used to constrain the model. Each of those parameters is varied between 0 and 25. For each combination of parameters, we generate 10 models, with different galaxies selected randomly as constraints in each one. The tested combinations of numbers of lensing constraints with and without spectroscopic redshifts  cover scenarios similar to the HFF clusters, including early pre-HFF models with a small number of constraints and as low as three spectroscopic redshifts, to the richest post-HFF data sets with hundreds of multiple images and dozens of new spectroscopic redshifts. 

We compare the test models on a few metrics: image plane rms as a measure of predictability of multiple images, model-predicted redshifts of systems without spectroscopic redshifts, mass distribution, and magnification. The results are compared with the fiducial model, rather than the simulation, in order to separate the systematic error induced by the constraints from other potential sources of systematics. The systematic error between the fiducial model and simulation truth may be sensitive to the exact modeling algorithm and parameterization choice, which is beyond the scope of this paper, yet will be important to investigate. We refer the reader to \citet{Meneghetti:2016xe} for current work, comparing different lensing methods. 

While we are only investigating this effect on a single method, this study is applicable to all strong lensing methods. Despite the variety of lensing methods (i.e., parametric versus non-parametric, see below), all methods are using the same lensing evidence to infer the mass distribution. 

%==================================================================================
%  STRONG LENS MODELING
%==================================================================================
\section{Strong lens modeling}
\label{sec:slmodeling}

While diverse in computational algorithms, all strong gravitational lens modeling codes have a common goal: solve the lens equation 

\begin{equation}
\beta_{ij} = \theta_{ij} - \alpha_{ij}
\end{equation}

\noindent for the image plane positions $\theta_{ij}$ of each multiply imaged background source that map to location $\beta_{ij}$ in the source plane. The solution is the deflection field tensor $\alpha_{ij}$, which can be differentiated to solve for the projected surface mass density $\Sigma$,

\begin{equation}
\nabla_{ij}\alpha_{ij} = \frac{4\pi G}{c^2}\frac{\dls(\zl,\zs)}{\ds(\zs)} \dl(\zl)\ \Sigma,
\label{eqn:SMD}
\end{equation}

\noindent where $\ds$, $\dl$, and $\dls$ are the angular diameter distances between the observer and the source at redshift $\zs$, between the observer and the lens at redshift $\zl$, and between the lens and source, respectively. When the lensing mass is contained to a single plane at $\zl$, then the deflection angle scales only with the lensing fraction

\begin{equation}
\frac{\dls}{\ds}(\zs) = \frac{\alpha_{ij}(\zs)}{\alpha_{ij}(z\rightarrow\infty)},
\label{eqn:dlsds}
\end{equation}

\noindent which is a function of a single variable $\zs$. Thus, it is clear that both the image plane position $\theta_{ij}$ and the source redshift $\zs$ are needed to determine the full three-dimensional lensing geometry needed to constrain $\Sigma$.

Strong lens modeling can include other observable strong lensing effects that constrain other first- and second-order derivatives of the lensing potential, including time delays, magnification ratios between different images of the same source, and flexion (distortion of the background source). These methods can be implemented in lens modeling codes, but are more prone to systematic error from the uncertainty of the observations themselves (i.e., lack of measured time delays, microlensing effects on magnification ratios, need for high-quality imaging for shape measurements). Measuring a centroid of an image of a lensed galaxy is straight-forward and typically the error in position is smaller than the errors in lensing due to cosmic variance and structure along the line of sight \citep{Limousin:2007fk}.

There are generally two schools of thought for lens modeling codes: parametric and non-parametric methods. Parametric methods (e.g., Lenstool, \citealp{Jullo:2007lr}; glafic, \citealp{Oguri:2010gr}; gravlens, \citealp{Keeton:2001lr}; light-traces-mass, \citealp{Broadhurst:2005qy} and \citealp{Zitrin:2009qf}; GLEE, \citealp{Suyu:2012qf,Suyu:2010xy}) assume that the lensing potential and mass distribution of the lens can be expressed as a superposition of parameterized density distributions, and those parameters can be solved from the lensing evidence.  A common assumption for these models is that mass distribution follows that of the light (i.e., cluster member galaxies are assigned their own halos); however, this assumption can be applied more rigorously or loosely depending on the method. Non-parametric methods (e.g., SWunited, \citealp{Bradac:2008kx}; SaW lens, \citealp{Merten:2011fk}; LensPerfect, \citealp{Coe:2008mz}; WSLAP, \citealp{Diego:2005ye,Diego:2007qf}; Grale, \citealp{Liesenborgs:2010gf}) make no assumptions about the mass distribution and instead solve for each ``pixel" in the surface mass density of an adaptive grid with higher resolution near the constraints and peak of the density distribution. Some codes have been hybridized to include aspects of both parametric and non-parametric techniques \citep[ex., ][]{Jullo:2009ij}.

%==================================================================================
%   THE FIDUCIAL LENS MODEL
%==================================================================================
\section{The fiducial lens model}
\label{sec:lens_model}

\subsection{The simulated cluster Ares}
\label{sec:ares}

\begin{figure*}
\center
\includegraphics[width=0.9\textwidth]{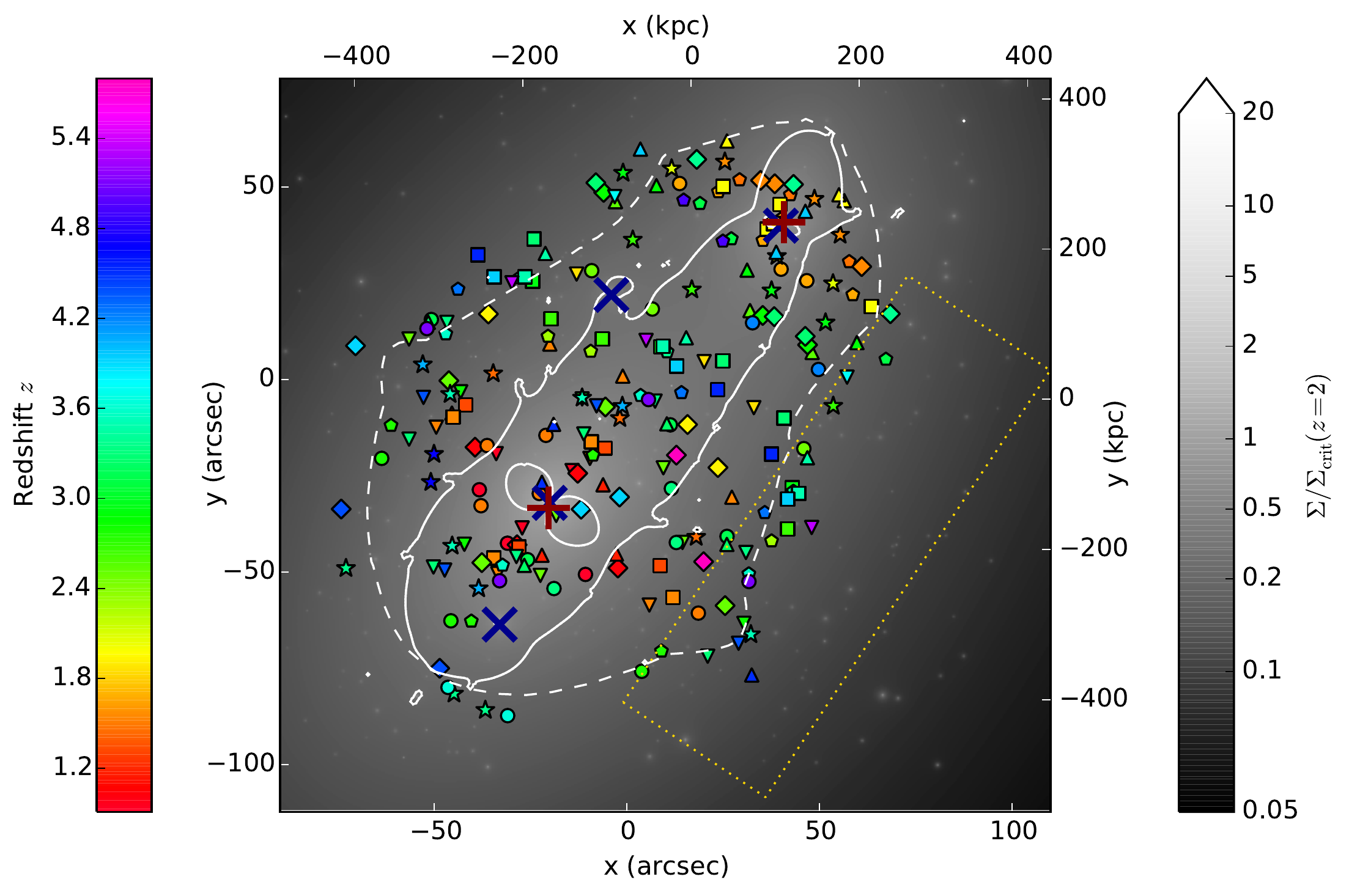}
\caption{Fiducial lens model of Ares. The grayscale image shows the projected surface mass density $\Sigma$ in units of the critical density $\Sigma_\mathrm{crit}=(c^2 \ds)/(4\pi G \dls \dl)$ at $z=2$. The locations of the multiple images used in the lens model are shown by the symbols, with colors indicating redshift. Images that match to the same source have the same redshift and are represented by the same symbol. The $z=2$ critical curve is shown by the solid white lines. The dashed white line indicates the region where the model predicts multiple images for $z=2$. The red crosses and blue x's mark the centers of the two cluster halos and four galaxy halos listed in \autoref{tab:params}, respectively. The gold dotted boxed region indicates the pixels used to generate the plots in Figures~\ref{fig:histograms}-\ref{fig:single_compare}.}
\label{fig:hst}
\end{figure*}

A full description of the Ares simulated cluster is given in \citet{Meneghetti:2016xe}. In short, the Ares cluster simulation is designed to mimic a massive cluster that is an efficient gravitational lens. The publicly available software MOKA \citep{Giocoli:2012lr} is used to create simulated lensing signals from clusters, including a number of scaling relations derived from $N$-body simulations, a mass-concentration relation, a subhalo mass function, and subhalo tidal stripping effects on truncation radii. The main cluster potential is composed of two triaxial clumps with masses $1.8\times10^{15}\ M_\odot$ and $1.3\times10^{15}\ M_\odot$ following the Navarro-Frenk-White profile \citep[NFW;][]{Navarro:1997qa}, separated by $\sim570$ kpc, along with two central cluster galaxies near the centers of these gravitational wells, which are modeled by triaxial profiles. The galaxy-scale halos are modeled as singular isothermal spheres.

The baryonic component of Ares follows a halo-occupation distribution (HOD) technique, where it is assumed that the stellar mass of a galaxy is tightly correlated with the depth of the gravitational potential well that it occupies. The B-band luminosities are assigned to each halo by MOKA using the relations described in \citet{Giocoli:2012lr}, which follow closely with the results by \cite{Wang:2006kk}. Then, a spectral energy distribution is assigned to each galaxy based on this luminosity and
empirical relations such as the morphology-density relation and the fraction of morphological types observed in clusters as a function of radius. Foreground galaxies and stars are added to the imaging; however, there is no additional mass along the line of sight to the cluster associated with these interlopers.

To simulate the lensed background Universe, the lensing signal from the cluster Ares is input into SkyLens \citep{Meneghetti:2010mi,Meneghetti:2008oz}, which ray traces real galaxies from the Hubble Ultra Deep Field \citep{Beckwith:2006rt} to the image plane. SkyLens creates mock \hst\ imaging of the results, which match the depth and wavelength coverage of the HFF observations.

The primary utility of Ares is for the ongoing HFF lens modeling comparison study \citep{Meneghetti:2016xe}. The same teams that modeled the HFF clusters, using different methods, were invited to compute lens models for Ares. Each team was given the simulated imaging along with a catalog of all the multiple image systems and redshifts, but were initially blind to the true mass distribution. The goal of this study is to identify how different methods reproduce the true mass and magnification of an HFF-like cluster when given identical inputs and to identify systematic errors across lens models that can be addressed when creating the best lens models. The initial ``blind" modeling took place in mid-2014, after which the mass and magnification were unveiled to the modeling teams. Our goals in this work can be thought of the tangent of those for the comparison study: rather than determine the systematics for different modeling methods using identical inputs, we are testing with a single method how varying the quantity and redshift information of constraints induces systematic errors.

\subsection{The lens model}
\label{subsec:fiducial_model}

We follow a similar methodology for modeling Ares as \citet{Johnson:2014tg}. We use the publicly available parametric modeling software Lenstool \citep{Jullo:2007lr}, which utilizes a Bayesian MCMC to explore the parameter space of the lensing distribution. We construct a lens model using all of the available lensing evidence that accurately reproduces the simulation mass and magnification. We refer to this model as the fiducial model, representing the best possible model we can create with our methods when all of the information (images, redshifts, mass distribution, etc.) is revealed. The fiducial model critical curves and images are shown in \autoref{fig:hst}.

\capstartfalse
\begin{deluxetable*}{cccccccc}
\tablecolumns{8}
\tablecaption{List of fiducial lens model constraints}
\tablehead{\colhead{} & \colhead{$x$ (\arcsec)} & \colhead{$y$ (\arcsec)} & \colhead{$e$} & \colhead{$\theta$ ($^\circ$)} & \colhead{$r_\mathrm{core}$ (kpc)} & \colhead{$r_\mathrm{cut}$ (kpc)} & \colhead{$\sigma_0$ ($\mathrm{km\ s^{-1}}$)}}
\startdata
cluster halo \#1 & $-20.3^{+0.1}_{-0.2}$ & $-33.3^{+0.1}_{-0.3}$ & $0.510^{+0.003}_{-0.016}$ & $50^\pm0.5$ & $100^{+3}_{-2}$ & [1500] & $1250^{+5}_{-7}$ \\[5pt]
cluster halo \#2 & $40.8^{+0.1}_{-0.3}$ & $40.9^{+0.4}_{-0.2}$ & $0.54^{+0.00}_{-0.05}$ & $74^{+1}_{-2}$ & $56\pm3$ & [1500] & $765^{+9}_{-5}$ \\[5pt]
galaxy halo \#1 & [-33.0] & [-63.6] & [0] & \nodata & [0] & $90^{+70}_{-0}$ & $240^{+130}_{-0}$ \\[5pt]
galaxy halo \#2 & [-20.0] & [-32] & $0.13^{+0.03}_{-0.07}$ & $156^{+27}_{-6}$ & [0] & [1500] & $502^{+10}_{-5}$ \\[5pt]
galaxy halo \#3 & [40.0] & [40.0] & $0.77^{+0.04}_{-0.20}$ & $5^{+5}_{-6}$ & [0] & [1500] & $290^{+6}_{-29}$ \\[5pt]
galaxy halo \#4 & [-4.0] & [22.0] & [0] & \nodata & [0] & [1500] & $213^{+4}_{-11}$ \\[5pt]
\hline \\[-5pt]
$L^\star$ galaxy & \multicolumn{4}{c}{$m_\star=20.00,\ z=0.5$ (ACS F606W)} & 0 & 20 & 100
\enddata
\tablecomments{The ellipticity is defined as $e=(a^2-b^2)/(a^2+b^2)$, where $a$ and $b$ are the semimajor and semiminor axes, respectively. The position angle is measure counterclockwise from the $+x$ axis. Parameters in square brackets are not optimized in the model. Errors represent the $1\sigma$ spread in values from the MCMC.}
\label{tab:params}
\end{deluxetable*}
\capstarttrue

\capstartfalse
\begin{deluxetable*}{cccc|cccc|cccc|cccc}{b}
\tablecolumns{16}
\tablecaption{List of lens model constraints}
\tablehead{\colhead{ID} & \colhead{$x$ (\arcsec)} & \colhead{$y$ (\arcsec)} & \colhead{$z$} &
		 \colhead{ID} & \colhead{$x$ (\arcsec)} & \colhead{$y$ (\arcsec)} & \colhead{$z$} &
		 \colhead{ID} & \colhead{$x$ (\arcsec)} & \colhead{$y$ (\arcsec)} & \colhead{$z$} &
		 \colhead{ID} & \colhead{$x$ (\arcsec)} & \colhead{$y$ (\arcsec)} & \colhead{$z$}}
\startdata
1.1 & -10.725 & -50.577 & 0.91 & 22.3 & -35.928 & 17.072 & 1.94 & 46.1 & 59.733 & 9.633 & 2.89 & 76.3 & 3.590 & -4.115 & 3.60 \\ 
1.2 & -30.979 & -42.475 & 0.91 & 23.1 & 56.449 & 46.502 & 1.96 & 46.2 & 31.230 & 28.455 & 2.89 & 76.4 & -46.980 & 11.896 & 3.60 \\ 
1.3 & -38.270 & -28.587 & 0.91 & 23.2 & 55.081 & 48.028 & 1.96 & 46.3 & 7.697 & 50.393 & 2.89 & 79.1 & -30.924 & -87.255 & 3.65 \\ 
2.1 & -27.147 & -38.528 & 0.93 & 23.3 & 26.002 & 62.007 & 1.96 & 47.1 & 42.968 & -28.046 & 2.92 & 79.2 & -46.404 & -79.916 & 3.65 \\ 
2.2 & -14.184 & -23.623 & 0.93 & 24.1 & 63.513 & 19.000 & 1.99 & 47.2 & -24.444 & 25.549 & 2.92 & 80.1 & 57.174 & 0.640 & 3.74 \\ 
2.3 & -33.898 & -19.225 & 0.93 & 24.2 & 36.400 & 39.062 & 1.99 & 47.3 & 8.798 & 8.540 & 2.92 & 80.2 & -3.132 & 47.515 & 3.74 \\ 
3.1 & -2.304 & -48.921 & 1.04 & 24.3 & 38.073 & 40.533 & 1.99 & 51.1 & 51.587 & 14.810 & 2.93 & 82.1 & -11.844 & -33.639 & 3.92 \\ 
3.2 & -28.476 & -42.877 & 1.04 & 24.4 & 39.757 & 45.470 & 1.99 & 51.2 & 37.575 & 23.124 & 2.93 & 82.2 & -70.416 & 8.812 & 3.92 \\ 
3.3 & -12.744 & -24.315 & 1.04 & 24.5 & 25.003 & 50.176 & 1.99 & 51.3 & -0.972 & 53.684 & 2.93 & 82.3 & -1.872 & -30.483 & 3.92 \\ 
3.4 & -39.384 & -17.531 & 1.04 & 25.1 & 53.529 & 24.966 & 2.09 & 54.1 & 67.260 & 5.295 & 3.11 & 85.1 & 38.782 & 33.107 & 3.95 \\ 
7.1 & -22.032 & -45.588 & 1.17 & 25.2 & 38.952 & 32.084 & 2.09 & 54.2 & 27.135 & 36.591 & 3.11 & 85.2 & 46.358 & 43.734 & 3.95 \\ 
7.2 & -2.700 & -45.269 & 1.17 & 25.3 & 11.642 & 54.807 & 2.09 & 54.3 & 18.989 & 45.759 & 3.11 & 85.3 & 3.533 & 59.881 & 3.95 \\ 
7.3 & -6.156 & -27.278 & 1.17 & 26.1 & -9.360 & 7.360 & 2.30 & 55.1 & -25.740 & -46.753 & 3.16 & 87.1 & 41.775 & -31.019 & 3.93 \\ 
7.4 & -45.396 & -8.665 & 1.17 & 26.2 & -20.448 & 11.256 & 2.30 & 55.2 & 26.011 & -40.700 & 3.16 & 87.2 & 12.927 & 3.545 & 3.93 \\ 
9.1 & 8.622 & -48.294 & 1.32 & 26.3 & 37.582 & -41.889 & 2.30 & 55.3 & 11.299 & -11.741 & 3.16 & 87.3 & -34.416 & 26.699 & 3.93 \\ 
9.2 & -28.008 & -43.477 & 1.32 & 28.1 & 45.962 & -17.912 & 2.46 & 55.4 & -50.724 & 15.617 & 3.16 & 90.1 & -38.448 & -54.228 & 4.05 \\ 
9.3 & -5.652 & -17.783 & 1.32 & 28.2 & 6.636 & 18.343 & 2.46 & 56.1 & 21.003 & -71.839 & 3.28 & 90.2 & -1.152 & -6.867 & 4.05 \\ 
9.4 & -41.832 & -6.569 & 1.32 & 28.3 & -9.108 & 28.311 & 2.46 & 56.2 & -50.184 & -48.671 & 3.28 & 90.3 & -52.956 & 3.924 & 4.05 \\ 
10.1 & 18.019 & -40.867 & 1.43 & 30.1 & -22.428 & -50.841 & 2.48 & 56.3 & -56.520 & -15.463 & 3.28 & 91.1 & 35.819 & -34.521 & 4.26 \\ 
10.2 & -1.800 & -10.009 & 1.43 & 30.2 & 14.455 & -42.302 & 2.48 & 56.4 & -11.160 & -14.100 & 3.28 & 91.2 & 14.194 & -3.374 & 4.26 \\ 
10.3 & -34.668 & 1.591 & 1.43 & 30.3 & 9.484 & -22.902 & 2.48 & 57.1 & 46.321 & 11.252 & 3.27 & 91.3 & -43.812 & 23.493 & 4.26 \\ 
12.1 & 57.741 & 30.644 & 1.48 & 30.4 & -56.520 & 10.542 & 2.48 & 57.2 & 38.274 & 16.363 & 3.27 & 93.1 & 49.749 & 2.646 & 4.20 \\ 
12.2 & 42.445 & 48.001 & 1.48 & 30.5 & -18.324 & -35.286 & 2.48 & 57.3 & -8.028 & 51.152 & 3.27 & 93.2 & 32.644 & 14.771 & 4.20 \\ 
12.3 & 29.281 & 51.904 & 1.48 & 32.1 & 25.509 & -58.685 & 2.51 & 58.1 & -26.604 & -48.205 & 3.24 & 94.1 & 29.016 & -68.471 & 4.38 \\ 
13.1 & 18.578 & -60.632 & 1.51 & 32.2 & -37.620 & -47.515 & 2.51 & 58.2 & 25.982 & -42.780 & 3.24 & 94.2 & -47.268 & -49.476 & 4.38 \\ 
13.2 & -37.836 & -32.729 & 1.51 & 32.3 & -46.188 & -0.270 & 2.51 & 58.3 & 10.403 & -11.463 & 3.24 & 94.3 & -52.776 & -4.618 & 4.38 \\ 
13.3 & -36.252 & -17.103 & 1.51 & 32.4 & -5.544 & -7.085 & 2.51 & 58.4 & -51.516 & 14.915 & 3.24 & 94.4 & -7.848 & -6.827 & 4.38 \\ 
13.4 & -20.988 & -14.466 & 1.51 & 34.1 & 48.111 & 7.079 & 2.54 & 60.1 & 40.775 & -9.977 & 3.27 & 97.1 & -48.564 & -74.958 & 4.39 \\ 
13.5 & -22.716 & -29.565 & 1.51 & 34.2 & 31.991 & 17.962 & 2.54 & 60.2 & -24.084 & 36.527 & 3.27 & 97.2 & -74.124 & -33.596 & 4.39 \\ 
14.1 & 5.861 & -58.540 & 1.53 & 34.3 & -2.952 & 46.247 & 2.54 & 60.3 & 24.957 & 4.897 & 3.27 & 102a.1 & -22.032 & -26.635 & 4.52 \\ 
14.2 & -33.480 & -50.367 & 1.53 & 35.1 & 41.776 & -38.743 & 2.66 & 61.1 & -36.712 & -85.809 & 3.36 & 102a.2 & -19.008 & -11.591 & 4.52 \\ 
14.3 & -9.576 & -20.461 & 1.53 & 35.2 & -6.372 & 10.565 & 2.66 & 61.2 & -44.860 & -81.572 & 3.36 & 102a.3 & 32.427 & -76.681 & 4.52 \\ 
14.4 & -49.464 & -12.354 & 1.53 & 35.3 & -19.692 & 15.798 & 2.66 & 61.3 & -72.889 & -48.979 & 3.36 & 102b.1 & -22.176 & -26.141 & 4.52 \\ 
15.1 & 60.981 & 29.367 & 1.55 & 37.1 & 53.595 & -6.837 & 2.68 & 62.1 & 43.239 & -29.010 & 3.39 & 102b.2 & -19.728 & -12.266 & 4.52 \\ 
15.2 & 38.407 & 50.882 & 1.55 & 37.2 & 16.880 & 23.387 & 2.68 & 62.2 & 10.455 & 7.209 & 3.39 & 102b.3 & 32.722 & -77.096 & 4.52 \\ 
15.3 & 34.726 & 51.760 & 1.55 & 37.3 & 1.557 & 36.319 & 2.68 & 62.3 & -27.792 & 26.473 & 3.39 & 102c.1 & -22.428 & -25.832 & 4.52 \\ 
16.1 & 27.280 & -30.507 & 1.54 & 38.1 & 8.997 & -70.498 & 2.75 & 63.1 & 12.854 & -42.291 & 3.33 & 102c.2 & -20.592 & -12.060 & 4.52 \\ 
16.2 & -1.044 & 0.941 & 1.54 & 38.2 & -40.356 & -62.721 & 2.75 & 63.2 & 11.577 & -28.277 & 3.33 & 102c.3 & 33.626 & -76.961 & 4.52 \\ 
16.3 & -19.980 & 9.256 & 1.54 & 38.3 & -8.784 & -19.603 & 2.75 & 63.3 & -18.900 & -54.243 & 3.33 & 108.1 & 37.588 & -19.356 & 4.55 \\ 
17.1 & 11.978 & -56.584 & 1.55 & 38.4 & -61.200 & -11.883 & 2.75 & 64.1 & -28.656 & -45.941 & 3.35 & 108.2 & -38.664 & 32.429 & 4.55 \\ 
17.2 & -34.524 & -46.299 & 1.55 & 40.1 & 3.877 & -75.744 & 2.78 & 64.2 & 30.928 & -44.889 & 3.35 & 108.3 & 23.592 & -2.581 & 4.55 \\ 
17.3 & -45.108 & -9.710 & 1.55 & 40.2 & -45.648 & -62.626 & 2.78 & 64.3 & 7.333 & -5.582 & 3.35 & 111.2 & -50.868 & -26.630 & 4.69 \\ 
17.4 & -9.216 & -16.172 & 1.55 & 40.3 & -63.612 & -20.433 & 2.78 & 64.4 & -46.692 & 14.850 & 3.35 & 111.3 & -50.040 & -19.338 & 4.69 \\ 
18.1 & 55.376 & 37.593 & 1.56 & 41a.1 & 30.397 & -63.300 & 2.83 & 66.1 & 68.410 & 17.123 & 3.38 & 114.1 & 24.941 & 35.967 & 4.94 \\ 
18.2 & 48.716 & 46.891 & 1.56 & 41a.2 & -42.156 & -42.774 & 2.83 & 66.2 & 43.273 & 50.699 & 3.38 & 114.2 & 14.758 & 46.620 & 4.94 \\ 
18.3 & 25.457 & 56.577 & 1.56 & 41a.3 & -11.592 & -5.008 & 2.83 & 66.3 & 18.154 & 57.203 & 3.38 & 115a.1 & -33.012 & -52.225 & 5.11 \\ 
19.1 & 58.666 & 22.058 & 1.66 & 41a.4 & -43.128 & -3.166 & 2.83 & 67.1 & 46.881 & -20.274 & 3.46 & 115a.2 & 31.712 & -52.424 & 5.11 \\ 
19.2 & 35.339 & 36.165 & 1.66 & 41b.1 & 30.179 & -63.441 & 2.83 & 67.2 & 15.427 & 10.887 & 3.46 & 115a.3 & 5.639 & -5.167 & 5.11 \\ 
19.3 & 23.792 & 48.810 & 1.66 & 41b.2 & -42.372 & -42.992 & 2.83 & 67.3 & -21.132 & 32.806 & 3.46 & 115a.4 & -51.876 & 13.269 & 5.11 \\ 
19.4 & 40.108 & 41.770 & 1.66 & 41b.3 & -11.484 & -5.232 & 2.83 & 68.1 & 44.641 & -29.530 & 3.49 & 115b.1 & -32.976 & -52.783 & 5.11 \\ 
20.1 & 46.730 & 25.702 & 1.67 & 41b.4 & -43.560 & -3.414 & 2.83 & 68.2 & 9.410 & 8.685 & 3.49 & 115b.2 & 31.031 & -52.778 & 5.11 \\ 
20.2 & 40.058 & 28.688 & 1.67 & 41c.1 & 30.615 & -63.139 & 2.83 & 68.3 & -26.424 & 26.682 & 3.49 & 115b.3 & 5.702 & -5.692 & 5.11 \\ 
20.3 & 13.711 & 50.941 & 1.67 & 41c.2 & -42.012 & -42.643 & 2.83 & 71.1 & 32.214 & -66.165 & 3.52 & 115b.4 & -52.524 & 13.017 & 5.11 \\ 
21.1 & 32.978 & -7.274 & 1.86 & 41c.3 & -11.664 & -4.763 & 2.83 & 71.2 & -45.324 & -43.159 & 3.52 & 117.1 & 47.994 & -38.406 & 5.34 \\ 
21.2 & -13.032 & 27.414 & 1.86 & 41c.4 & -42.804 & -2.962 & 2.83 & 71.3 & -45.864 & -3.768 & 3.52 & 117.2 & 5.000 & 10.195 & 5.34 \\ 
21.3 & 20.084 & 4.657 & 1.86 & 45.1 & 46.930 & 9.042 & 2.85 & 71.4 & -11.592 & -4.703 & 3.52 & 117.3 & -29.808 & 25.127 & 5.34 \\ 
22.1 & 23.659 & -22.823 & 1.94 & 45.2 & 35.297 & 16.677 & 2.85 & 76.1 & 31.673 & -50.445 & 3.60 & 122.1 & 19.933 & -47.271 & 5.80 \\ 
22.2 & 15.774 & -11.652 & 1.94 & 45.3 & -6.012 & 48.638 & 2.85 & 76.2 & -32.256 & -48.181 & 3.60 & 122.2 & 12.879 & -19.594 & 5.80

\enddata
\tablenotetext{}{Coordinates are relative to the origin defined in \autoref{fig:hst}.}
\label{tab:constraints}
\end{deluxetable*}
\capstarttrue

The mass distribution is parameterized by pseudo-isothermal elliptical mass distributions \citep[PIEMD or dPIE; ][]{Limousin:2005cr}; the profile is described by a fiducial velocity dispersion $\sigma_0$ to normalize the potential, an ellipticity and position angle, and core radius $r_\mathrm{core}$ and cut radius $r_\mathrm{cut}$, which control the inner and outer slopes of the profile, respectively. A summary of those halo parameters is given in \autoref{tab:params}. We use two halos to represent the dark matter cores in the cluster, which were also included in our ``blind" lens model. We include the masses of galaxy cluster members as small perturbers to the smooth dark matter potential of the cluster. The galaxies are selected by red-sequence membership and their halo parameters are scaled by their brightness following

\begin{eqnarray}
\sigma_0&=& \sigma_0^\star \left(\frac{L}{L^\star}\right)^{1/4}, \nonumber \\
r_\mathrm{core} &=& 0 \\
r_\mathrm{cut} &=& r_\mathrm{cut}^\star \left(\frac{L}{L^\star}\right)^{1/2} \nonumber
\end{eqnarray}

\noindent where $\sigma_0^\star, r_\mathrm{core}^\star, r_\mathrm{cut}^\star$ are the parameters of an $L^\star$ galaxy at the simulated cluster redshift $z=0.5$. For four cluster galaxies (including the two brightest galaxies in both cores), we allow some of the parameters to deviate from the scaling relations and to be guided by the lensing of nearby multiple images, as we routinely do in lens models of real clusters. Galaxy halo \#1 (as indicated in \autoref{tab:params}) is a massive galaxy that has a significant impact on the location of the critical curve in the southwestern portion of the image plane. Galaxy halos \#2 and \#3 lie near the centers of the two massive cluster halos and thus have an impact on the locations of the radial arcs across the entire cluster. The fourth galaxy halo is massive enough to produce its own protrusion of the tangential critical curve created by the two massive cluster halos, thus influencing the lensing of several nearby images. These four galaxies are needed for models with many constraints; however, their parameters are more difficult to constrain in the case when there are no images within a few arcseconds of the halo.

Our treatment of the scaling relations for Ares deviate from those of the models in \citet{Johnson:2014tg} due to the construction of the simulation. We did not include shape information from the light distribution to guide the galaxy shapes in the lens model and instead modeled the halos as single isothermal spheres ($\mathrm{ellipticity} = 0$, $r_\mathrm{core} = 0$) in order to match closely to the parameterization of the simulation. We ran a simple optimization to explore which scaling parameters produce a close match between the simulated halos and lens model halos. All three of these parameters are highly degenerate when determining the mass of a halo and thus no single parameter combination was determined to be a significantly better fit over the others. Additionally, the typical scale of $r_\mathrm{cut}$ is several tens to 100 kiloparsecs; at this scale the halo of a galaxy begins to overlap with neighboring galaxies and the main cluster potential starts to dominate the local surface mass density. Thus, $r_\mathrm{cut}$ is difficult to constrain. We selected a parameter combination that was reasonable with those of previous strong lensing models and matched well with the simulation: $\sigma_0^\star=100\ \mathrm{km\ s^{-1}}$, $r_\mathrm{cut} = 20$ kpc, and $m_\mathrm{F606W}^\star=20.0$. Due to a different choice of scaling relations and photometric band used for scaling, we do not expect the simulation and lens model to match perfectly. Our goal with this optimization is to minimize the effects of scaling parameter selection on the overall systematic errors of the lens model we are attempting to measure. The mass of the fiducial model is reconstructed with an accuracy of $-0.24^{+0.23}_{-0.30}\%$ of the simulation mass within 500 kpc of the cluster center.

A list of simulated multiple images, their locations, and their redshifts was released along with the simulated data. We altered this list to comply more closely with one that would have been created by a lens modeler identifying images by eye. We made slight adjustments ($<0\farcs1$) to the location of the image constraints to match the same features of a galaxy in all its multiple images. Additionally, for three image systems with more extended sources, we include multiple positional constraints corresponding to different unique features within the lensed galaxy. Finally, we purged the list of images that would not be detectable if the search was done by eye (for example, images behind a large galaxy, too faint to be visible), such that our identification quality matched that of deep HFF-based lens models. \autoref{fig:hst} shows the locations and redshifts of our final list of 232 multiple images from 66 unique sources ($\Nfid=66$) with $0.91<z<5.80$ (and are listed in \autoref{tab:constraints}). The image plane rms for the fiducial lens model of Ares for all 66 image systems is $0\farcs58$ (see \S\ref{subsec:rms} for definition and further discussion), which is on par with the scatter quoted for Lenstool-based models of the HFF clusters \citep{Jauzac:2014qd,Jauzac:2015xy,Jauzac:2016dn,Johnson:2014tg,Richard:2014gf,Sharon:2015rz,Treu:2016lr}. 

%==================================================================================
%   METHODOLOGY
%==================================================================================
\section{Test Models}
\label{sec:methods}

The fiducial model of Ares represents an idealized scenario for creating a lens model, one where many image systems are known with certainty and all systems have confirmed redshifts. However, this scenario would be considered extreme compared to models of real clusters, which typically have fewer multiple image systems and even fewer spectroscopic redshifts. To represent types of cluster lens models that currently exist, we create new models of Ares using ``jacknifed" subsets of images from the full list of image systems. We randomly select $n=0,5,10,15,20,25$ image systems with their known (spectroscopic) redshifts and $m=0,5,10,15,20,25$ image systems with unknown redshifts and remodel the cluster with a total number of image systems $N=n+m<\Nfid$. For the $m$ systems without known redshifts, we only include image positions as constraints in the model and leave redshift as a free parameter with a uniform random prior probability distribution function range of $0.6<z<7$ (see \S~\ref{subsec:photoz} for a treatment/discussion of photometric redshifts). We run 10 different models for each combination of $n,m$ for better statistics, each with a unique set of images. We refer to these models with different $n,m$ as the ``test models" from here-on.

We choose to run the test models with the same parameterization as the fiducial model (i.e., same free and fixed parameters and priors as \autoref{tab:params}) so that we can directly compare these models with the fiducial model. It is true that models with lower $N$ may not be able to constrain all of the free parameters of the fiducial model. By basing the parameterization of the test models off of the fiducial model, we are including some knowledge a priori about the mass distribution for which a given set of $N$ images alone may not be able to provide enough evidence (i.e., existence of a secondary halo, shape of central galaxies, etc.). In a truly blind scenario, it is likely that a lens modeler would choose a different parameterization; however, the choice of parameterization on a model-by-model basis is not easy to simulate. With this caveat in mind, the systematic errors for smaller $N$ stated here are likely lower limits that do not reflect the choice in parameterization as a function of $N$.

Lens modeling is a computationally intensive task, as proper modeling in the image plane entails inverting the lens equation and computing the scatter for each multiple image, which requires scanning many image plane pixels for matching source plane positions. The newest versions of Lenstool (v6.7 and above) have built-in parallelization that dramatically reduces computation time; however, a model can take days to weeks to run under optimal parallelization. Since we ran all of the models for this work with image plane optimization, the computation time is considerable. We used the Flux High Performance Cluster at the University of Michigan to compute these models, using Lenstool version 6.8 on eight nodes with 20 core processors (two 10-core 2.8 GHz Intel Xeon E5-2680v2 processors) and 96 GB RAM over the course of four months, where all 350 test models ran continuously in queue. In order to increase the number of models running in parallel on a single node, the models with fewer total image systems $N$ were assigned to run below node capacity, such that the $N_{cores}=\mathrm{floor}(N/2)+1$ and up to 20 cores. Each model was run with the Lenstool parameter for Bayesian rate set to the maximum of 0.5 and for only a set of 5010 models in the MCMC. The total wall time for the test models clocks in at $\sim329,000$ core-hours.

%==================================================================================
%   RESULTS
%==================================================================================
\section{Results}
\label{sec:results}

The different combinations of $n$ (spectroscopic redshifts) and $m$ (free parameter / unknown redshifts) result in 35 model families. For better statistics, each of these combinations was sampled 10 times, for a total of 350 models. We now compare the lensing outputs of these 35 model families against each other and against the fiducial lens model.  Below, we investigate the dependence of several diagnostics on the total number of lensed galaxies used as constraints ($N=n+m$), the number of spectroscopic redshifts ($n$), and the fraction of spectroscopic redshifts ($n/N$). 

\subsection{Image predictability}
\label{subsec:rms}

The image plane rms scatter of multiple image systems is a measure of how accurately a lens model can reproduce the locations of images. It is effectively the quantity that is being minimized during image plane optimization (the $\chi^2$ is the image plane rms normalized by the estimated error in image position). The locations of multiple images are transformed to the source plane and then relensed to other locations in the image plane and the scatter is computed based on the separation of the predicted and observed locations.

\begin{figure*}
\center
\includegraphics[width=\textwidth]{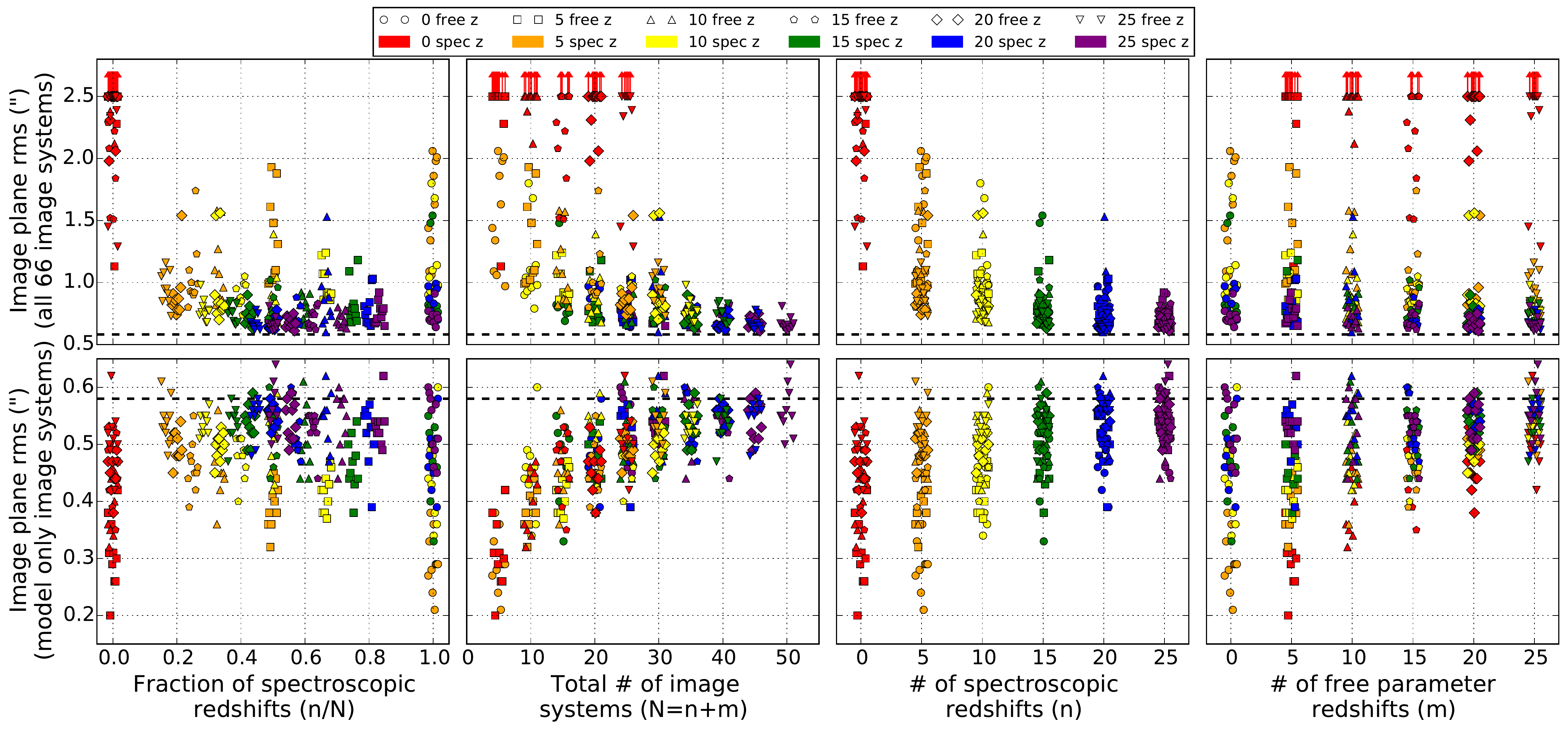}
\caption{Image plane rms for all the test models plotted vs. the fraction of spectroscopic redshifts, the total number of image systems, the number of spectroscopic redshifts, and the number of free parameter redshifts. The colors and shapes of the points represent the number of spectroscopic redshifts and the number of free parameter redshifts used in the model, respectively (see legend at the top). The top panels show the image plane rms values computed for all 66 image systems using the true redshift values. The bottom panels show the image plane rms values computed only from the images used as constraints in the lens model and model-predicted redshifts.  The dashed line indicates the value of the image plane rms for the fiducial model and is computed from all 66 image systems and true redshifts.}
\label{fig:rms}
\end{figure*}

To see the effects of adding image systems with spectroscopic or unknown redshifts, we compute the image plane rms using all 66 image systems and their true redshifts for each of the test models in \autoref{fig:rms} (top panels). This test shows how well the model can reproduce lensing in many parts of the image plane, not only where constraints are located. We find that increasing the total number of systems $N$ decreases the rms scatter asymptotically toward the fiducial model rms. We also find that this trend is true for increasing number of spectroscopic redshifts $n$ and only weakly for increasing free parameter redshifts $m$ for models with $n<10$. 
Models will improve significantly in image predicting power when more image systems are included in the model, especially those with spectroscopic redshifts. However, this effect plateaus for models with either $N>25$ or $n>20$, when the exact selection of the constraints rather than quantity determines the level of systematic error in image plane rms. In clusters with many lensed galaxies, modelers often rely on preliminary lens models in order to predict the locations and identify new sets of lensed images. This result shows the importance of having spectroscopic redshifts in these preliminary models, as at least $n>10$ spectroscopic redshifts are needed in order to robustly distinguish between multiple image candidates based on their model-predicted location ($\mathrm{rms}<1\farcs0$). In particular, all models with no spectroscopic redshifts have poor rms.

It is worth emphasizing that the image plane rms we computed using all 66 image systems would not be the value quoted for a typical lens model of a real cluster in the literature. In reality, modelers compute the rms only for the image systems used in the model and use the redshift solutions from the best-fit model for the systems without spectroscopic redshifts (not the true redshifts, as these are not known). To demonstrate this discrepancy, we plot the rms value computed using only the image systems and the model-derived redshifts\footnote{This image plane rms corresponds to the value from the output file in the Lenstool software for the best-fit model.} in \autoref{fig:rms} (bottom panels). We see that this value tends to be much lower than the fiducial value, and the trends for this rms are the reverse of the top panels. The rms value computed in the bottom panels is a measure of goodness of fit, adding more free parameter redshifts increases model flexibility and adding more spectroscopic systems increases the number of constraints without increasing the number of free parameters. Models that are less flexible with more constraints produce higher rms values in the bottom panels, indicating a worse model fit; however, these models are better at predicting the locations of images across the entire image plane, as indicated by the rms in the top panels.

\autoref{fig:rms} demonstrates the need for caution when relying on the image plane rms value to judge lens model fidelity, especially when many image systems with unknown redshifts are included in the model. Since the deflection field scales with distance to the source, the image plane rms will depend on the redshift of the source. For spectroscopic systems, the redshift is fixed; however, the free parameter redshift of image systems included in the model, by the construction of a maximum likelihood optimization, will take on a value for the redshift that helps to minimize the overall image plane rms, which may or may not coincide with the correct redshift. While models without free parameter redshifts have flexibility and report low image plane rms, they have the potential to encounter parameter degeneracies between the mass distribution and source plane redshifts.

\begin{figure*}
\includegraphics[width=\textwidth]{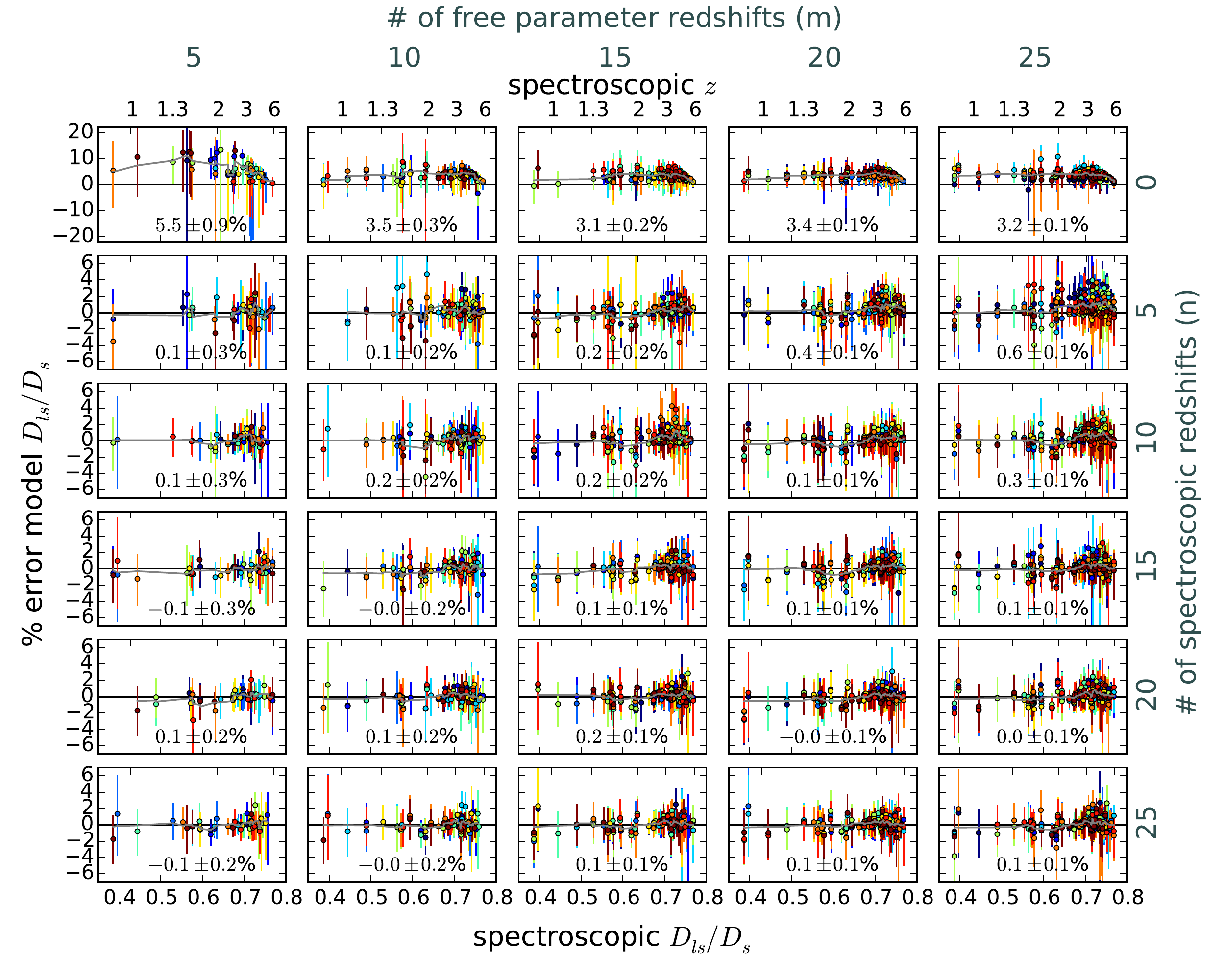}
\caption{Error in best-fit model redshift parameter of image systems used as constraints in sets of 10 models using different numbers of spectroscopic redshifts and free parameter redshifts. The redshifts are plotted in terms of the lensing fraction $\dls/\ds$, a function of source redshift that scales the deflection angle of the lens. The error bars represent the $1\sigma$ errors computed from the MCMC chain. The colors match the free parameter redshifts used in the same model. The gray line indicates the rolling average across $\dls/\ds$ from all models. The value at the bottom of each is the weighted mean error in the lensing fraction for all models. Note that the top row ordinates have a different scale from the other plots.}
\label{fig:dlsds}
\end{figure*}

\subsection{Model-predicted Redshifts}

We investigate how accurately models using free parameter redshifts predict the true redshift of those image systems. \autoref{fig:dlsds} shows the error between model-derived redshift and the true redshift of the system. We plot these errors in terms of the lensing fraction $\dls/\ds$ rather than $\zs$; as shown in \autoref{eqn:dlsds}, the deflection angle tensor $\alpha_{ij}$ scales linearly with the lensing fraction. The typical error in model predicted $\dls/\ds$ is $<2\%$ in all cases and tends to be lower for models that have higher fractions of image systems with spectroscopic redshifts. Interestingly, models with low fractions of image systems that have spectroscopic redshifts tend to predict redshift solutions that are more often biased to higher values for systems with  $z>2$. Nearly all of the model-predicted $\dls/\ds$ of models with $n=0$ are biased high by 5-10\%.

\subsection{Mass}
\begin{figure*}
\center
\includegraphics[width=\textwidth]{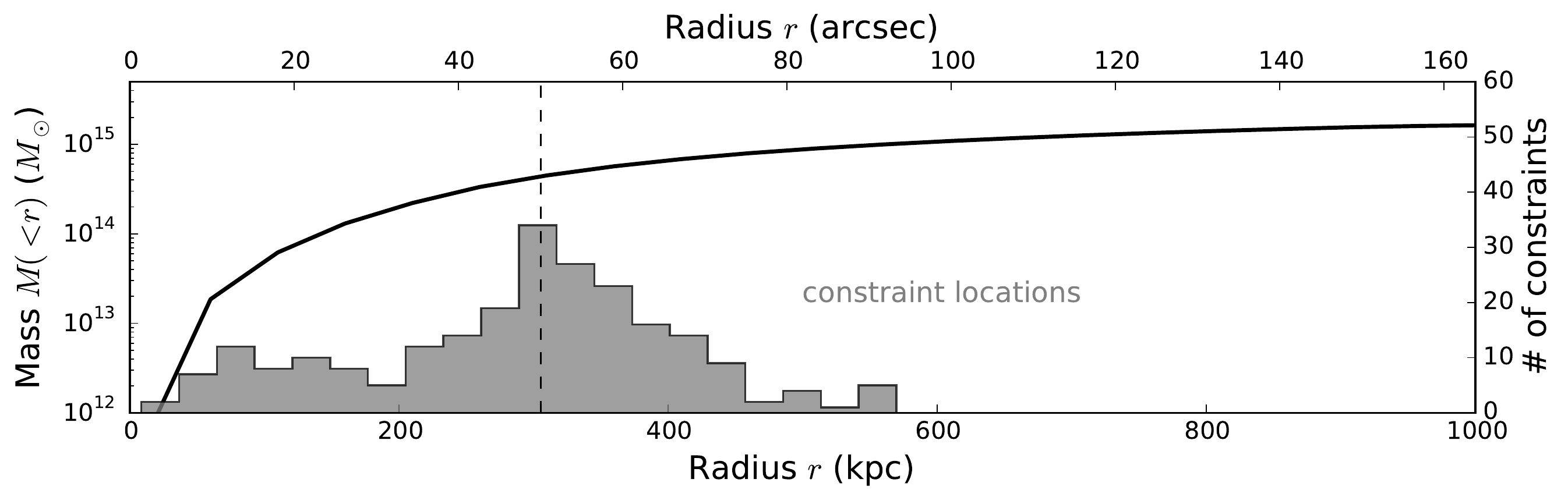}
\includegraphics[width=\textwidth]{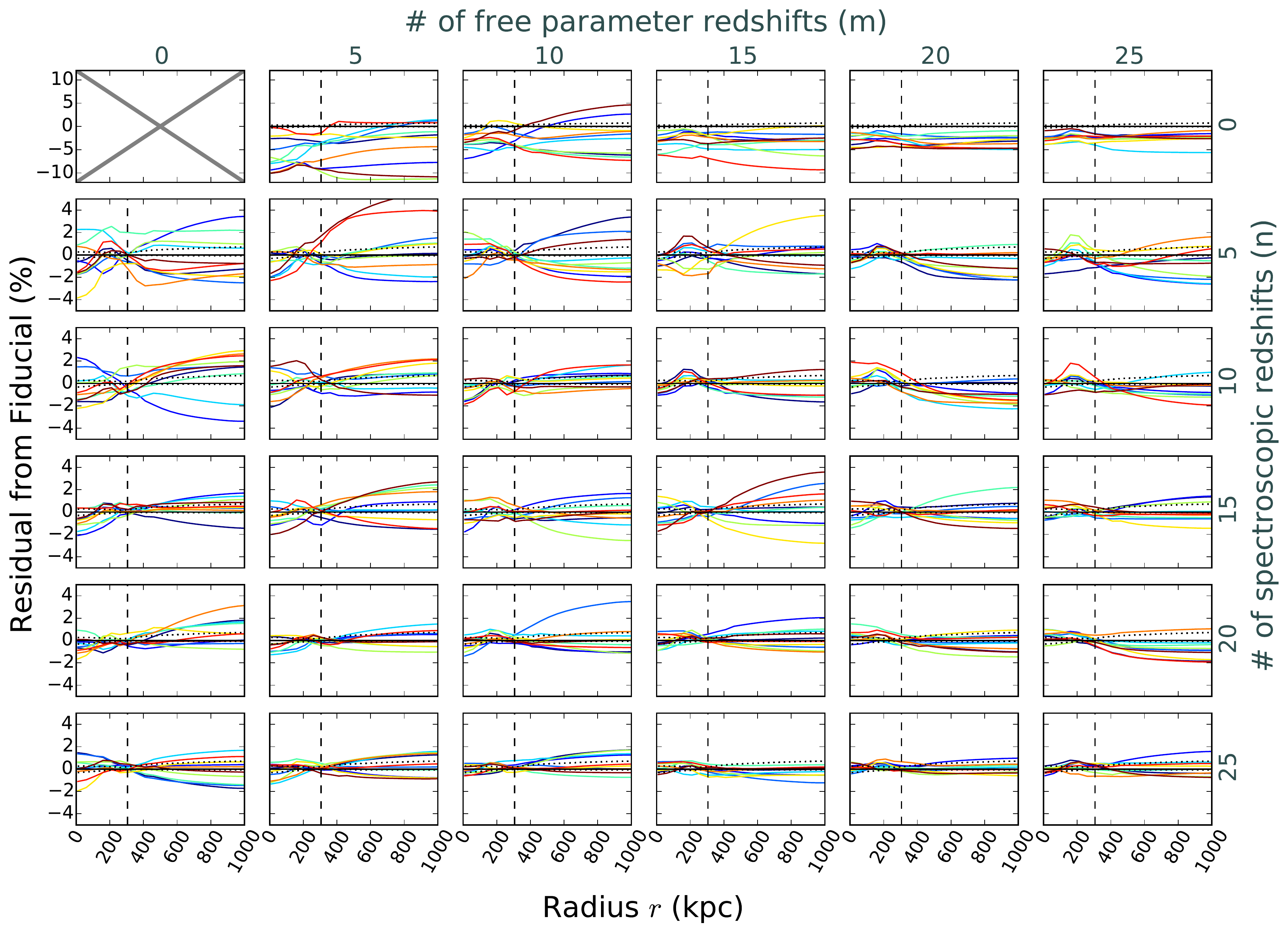}
\caption{(Top) radial mass profile for the fiducial model. The histogram shows the projected radii of all the constraints used in the model. The dashed vertical line is the median projected radius of the arcs at $r_\mathrm{arcs}=305$ kpc. (Bottom) Radial mass profile residuals from the fiducial model for all test models with different numbers of spectroscopic and free parameter redshifts used as constraints. The dotted lines represent the $1\sigma$ statistical error in the fiducial model mass profile estimated from the MCMC. The dashed vertical line matches $r_\mathrm{arcs}$ from the top plot. Note: the top row ordinates have a different scale from the other plots.}
\label{fig:mass}
\end{figure*}

In \autoref{fig:mass}, we plot the projected mass profile of the cluster for the fiducial model (top) and residual from the fiducial model for all of the test models (bottom). We find that the errors in the enclosed mass are typically $<4\%$ out to 1 Mpc for models with $n>0$. Models with $n=0$ are generally biased toward lower masses, which is consistent with the model predicting higher redshifts for the free parameter image systems. For models with $n>0$, the errors are generally lowest at radii around the ``arc radius," $r_\mathrm{arcs}=305$ kpc, defined as the median image plane projected distance of images used as constraints in the fiducial model; it is comparable to the formal definition of the Einstein radius. Test model combinations with at least five spectroscopic redshifts have errors of $<1\%$ around the arc radius.

\subsection{Magnification}
\label{subsec:magnification}
The magnification describes the amplification of the solid angle of a lensed object from source plane to image plane and is derived from the lensing Jacobian tensor

\begin{equation}
A_{ij}\equiv \frac{\partial\beta_{ij}}{\partial\theta_{ij}}=\delta_{ij}-\frac{\partial\alpha_{i}}{\partial\theta_{j}},
\end{equation}

\noindent which describes the translation from source plane $\beta$ to image plane $\theta$. The magnification $\mu$ is the inverse determinant of this tensor,

\begin{equation}
\mu = \frac{1}{|\det A_{ij}|},
\end{equation}

\noindent which becomes a nonlinear combination of first-order derivatives of $\alpha$. Based on its complexity, we expect the magnification factor to have more complicated relations with observable quantities than the diagnostics discussed earlier.

\begin{figure*}
\center
\includegraphics[width=1.0\textwidth]{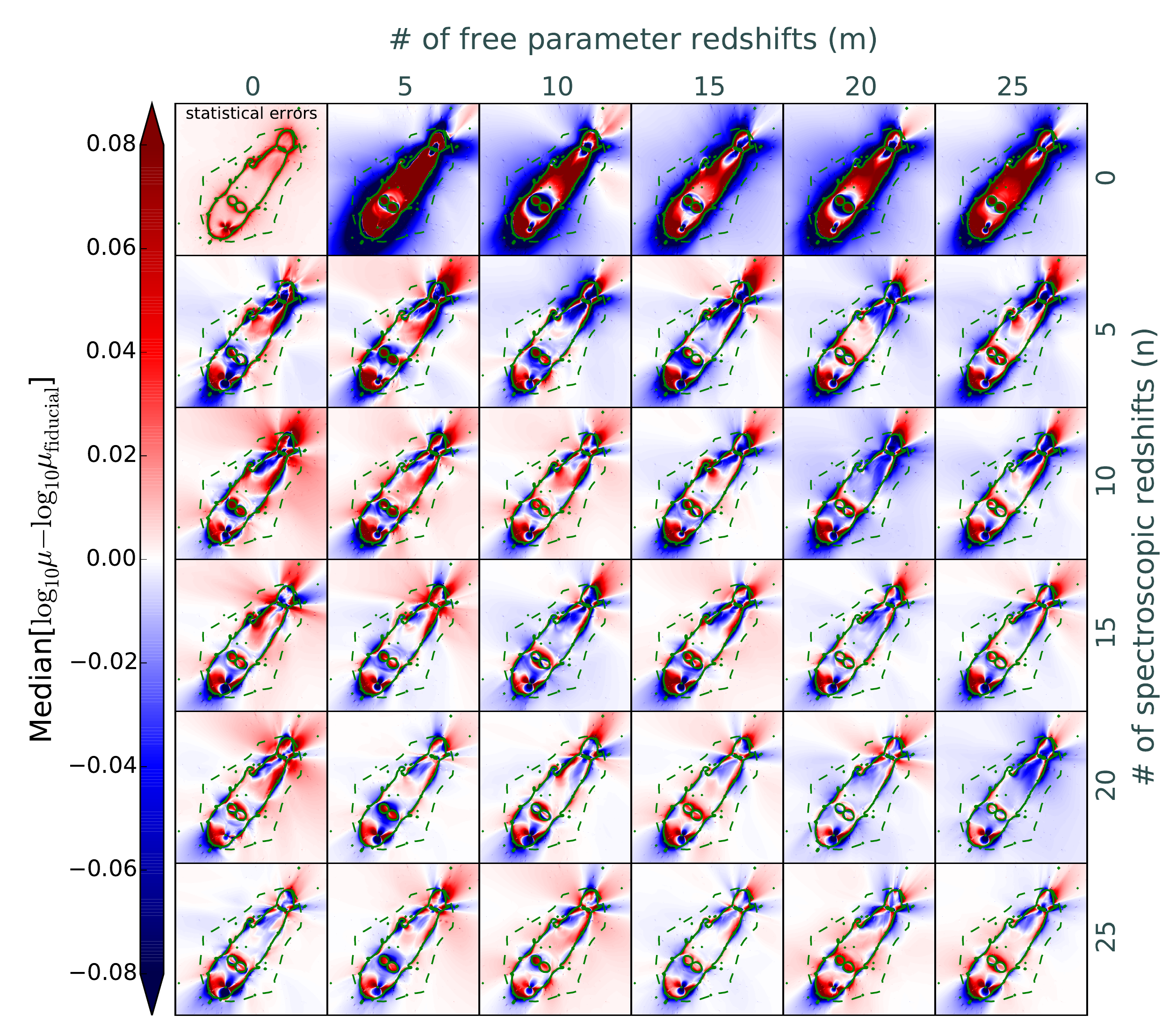}
\caption{Median error in the magnification maps for $z=2$ for each set of models with various numbers of constraints with spectroscopic redshifts and free parameter redshifts. The error is computed with respect to the magnification map of the fiducial model. The $z=2$ critical curve for the fiducial model is shown in solid green and the region enclosed by the dashed green line is the extent of image multiplicity for sources at $z=2$. The top-left panel shows the statistical errors in magnification for each pixel of the fiducial model. Each panel has dimensions of $200\times200$\arcsec.}
\label{fig:magnification_bias}
\end{figure*}

In \autoref{fig:magnification_bias}, we plot the error in magnification at each image plane position corresponding to a source at $z=2$ for each test model combination (i.e., median magnification of each pixel across all models with same $n,m$) relative to the fiducial model magnification. These maps effectively show the bias in magnification when selecting $n,m$. We choose to display $z=2$ because it corresponds to a middle value of $\dls/\ds$ for all the sources used as constraints.

For models with $n>0$, the magnification errors are all quite similar. Across all models, the magnification is most accurate in regions of lower magnification ($\mu<10$) and along the straight portion of the critical curve, the region where most of the multiple images are located. A straight critical curve implies that the vector of the deflection angle is nearly constant in terms of direction and only changes in amplitude; solving the lens equation in this region becomes one-dimensional. At the high-curvature portions of the critical curve, the tangential shear is strong and the deflection angle changes rapidly in both amplitude and direction. Also, objects here are highly magnified, but their image multiplicity becomes unity. These singly imaged sources are indeed strongly lensed, but are not used as strong lensing constraints for this modeling method. Some methods can use single images as constraints; however, doing so greatly increases computing time in order to reject models producing multiple images. Additionally, the flexion of these highly magnified single-image systems could be included in modeling methods to better constrain the mass distribution where there are no multiple images \citep[see][]{Cain:2011ab}.

\begin{figure*}
\center
\includegraphics[width=1.0\textwidth]{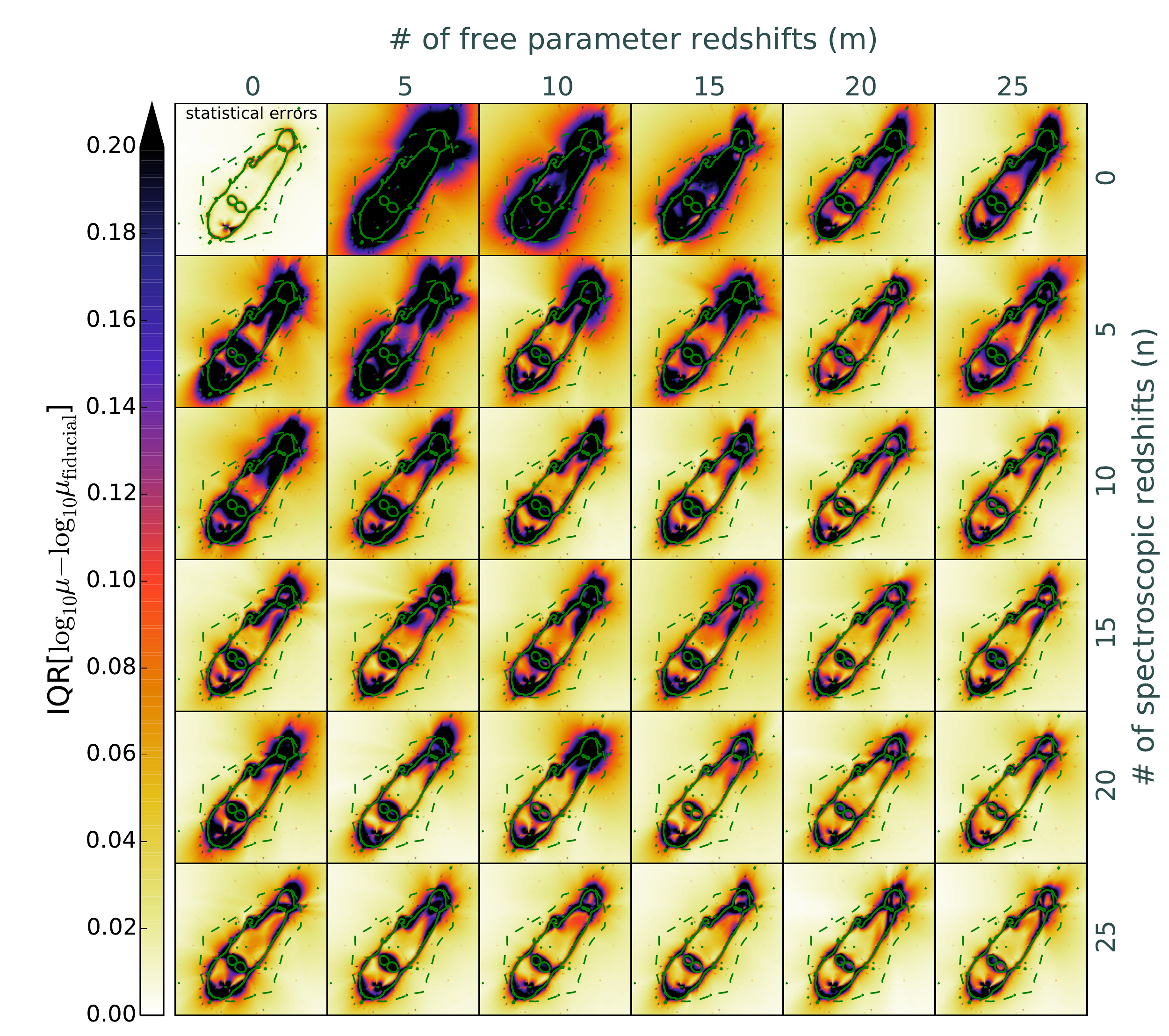}
\caption{Interquartile range (IQR) of errors in the magnification maps for $z=2$ for each set of models with various numbers of constraints with spectroscopic redshifts and free parameter redshifts. The error is computed with respect to the magnification map of the fiducial model. The $z=2$ critical curve for the fiducial model is shown in solid green and the region enclosed by the dashed green line is the extent of image multiplicity for sources at $z=2$. The top-left panel shows the statistical error range in magnification for each pixel of the fiducial model. Each panel has dimensions of $200\times200$\arcsec.}
\label{fig:magnification_spread}
\end{figure*}

While \autoref{fig:magnification_bias} shows the bias of the magnification for all regions in the image plane over a slew of different models, \autoref{fig:magnification_spread} shows the interquartile range (IQR\footnote{We define the IQR as the difference between the 75th and 25th percentile models. These percentiles correspond to the average magnification between the 8th/9th-ranked and 2nd/3rd-ranked models, respectively.}) in magnification, i.e., how consistent the systematic errors in magnification are relative to the fiducial models when different sets of constraints are used for the same $n,m$. We plot the IQR to eliminate the effects of potential outlying models in our analysis. We see similar trends to those of \autoref{fig:magnification_bias}: the IQR in magnification between models is lower for regions with low magnification and along the straight portion of the critical curve. We see a very clear trend with reduced spread in magnification error throughout most of the image plane with higher $N$.

\begin{figure*}
\center
\includegraphics[width=1.0\textwidth]{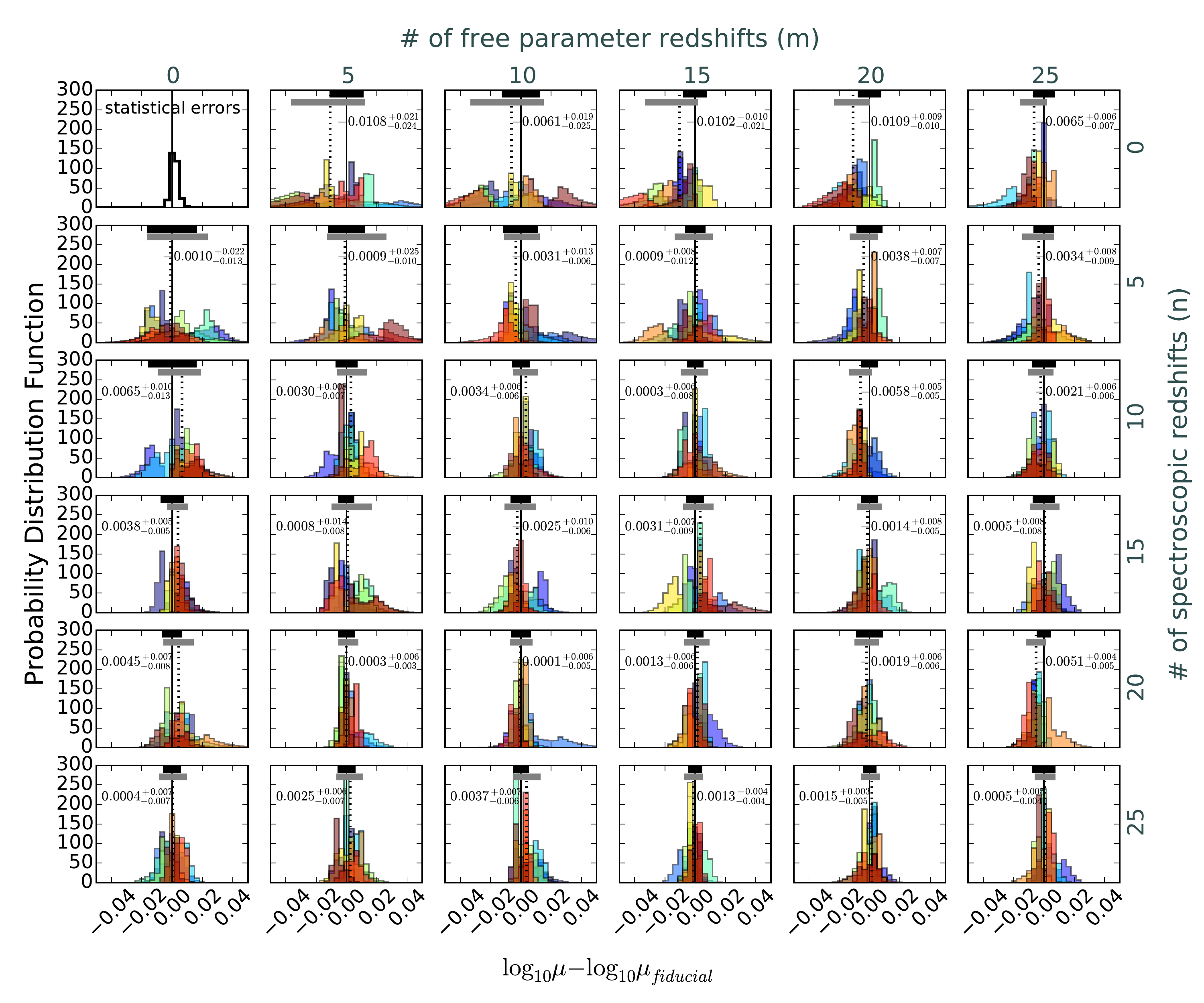}
\caption{Histograms of magnification error ($z=2$) for the region of pixels shown in \autoref{fig:hst} for models with different numbers of spectroscopic and free parameter redshifts. Each color shade represents a unique model constructed with different random subsets of images used as constraints. The top-left panel shows the $1\sigma$ statistical errors of each pixel in the fiducial model. The black bar on top shows the typical $1\sigma$ statistical error for a test model. The dashed vertical line and horizontal gray bars show the median and $1\sigma$ range in magnification error distribution of all test models combined, and these values are displayed in each panel.}
\label{fig:histograms}
\end{figure*}

We attempt to quantify the systematic errors in \autoref{fig:histograms} by looking at the distribution of magnification errors across the image plane. We create histograms of the magnification error for each pixel for each individual test model. We only examine pixels located in a rectangular region with bounds selected arbitrarily such that it lies in the lower right $(+x,-y)$, aligned roughly with the critical curve of the cluster, to avoid pixels near the curved portion of the critical curve. This region of pixels is shown in \autoref{fig:hst}. We also only select pixels with $\mu_\mathrm{fiducial}<20$ to avoid high magnifications induced locally by cluster member galaxies. We see that models with lower $N$ tend to produce magnifications that are typically biased low; however, beyond $N\geq25$, the distributions of models appear to be similar and with negligible bias, with a typical error of about 2\%.

It is noticeable across all test models that the variation in the distribution of magnifications is quite significant for the low total number of image systems, even amongst test models with identical $n,m$. This indicates that it is not necessarily the quantity of image systems or redshifts, but rather the selection of these constraints that drives systematic error. We examined closely a few of the models with outlying distributions in \autoref{fig:histograms} and found that the random selection of spectroscopic redshift systems for those models was either unevenly distributed spatially in the image plane or unevenly distributed in redshift space.

\begin{figure*}
\center
\includegraphics[width=\textwidth]{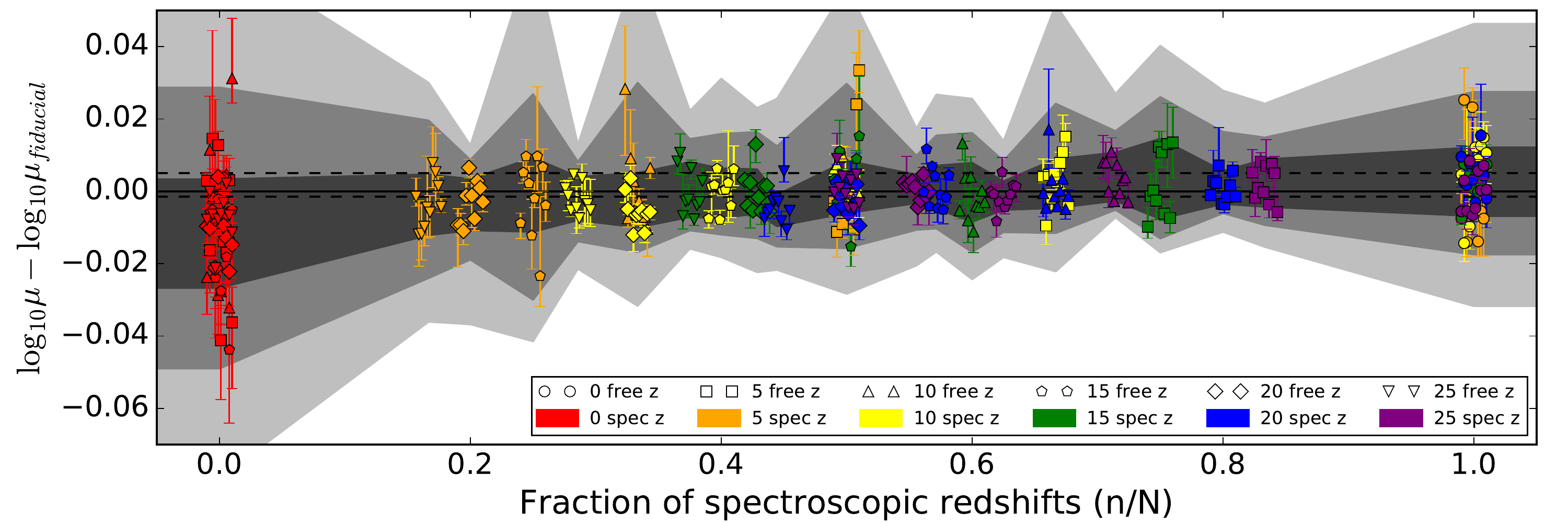}
\includegraphics[width=\textwidth]{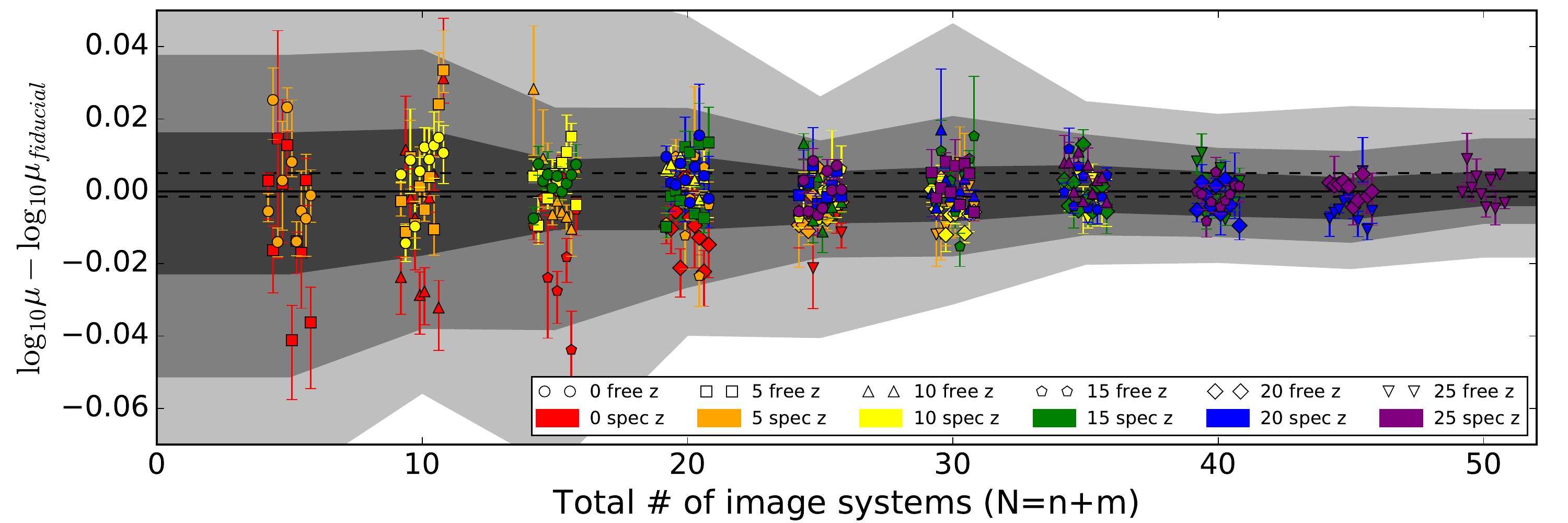}
\caption{Relative magnification error ($z=2$) for the region of pixels shown in \autoref{fig:hst} for all the test models vs. the fraction of spectroscopic redshifts $n/N$ (top) and the total number of image systems $N$ (bottom). The values are a median and $1\sigma$ range of values within the region for the best-fit models. The different shapes and colors indicate the number of free parameter redshifts and spectroscopic redshifts used in the model, respectively. The dashed lines indicated the $1\sigma$ statistical errors for the fiducial model. The gray contours represent the 1, 2, and 3$\sigma$ ranges for each block of the test models in fraction/number of systems. Note that the abscissa values for the test models within each grouping of test models have been slightly offset horizontally for display purposes. There is a clear trend of improved magnification error with the total number of images, but no dependence on the fraction of spectroscopic redshifts.}
\label{fig:specz_fraction}
\end{figure*}

In \autoref{fig:specz_fraction}, we plot the relative magnification error and spread for all test models (defined as they are in \autoref{fig:histograms}) versus the fraction of spectroscopic redshifts $n/N$ and total number of image systems $N$. We report no clear trend in magnification error or spread with spectroscopic redshift fraction, except that models with no spectroscopic redshift are biased toward lower magnifications and have a $1\sigma$ spread of about 3\%.  There is a clear trend in decreasing systematic error with $N$, and for $N\geq25$, the $1\sigma$ magnification error stays constant at about 1\%.

%==================================================================================
%   DISCUSSION
%==================================================================================
\section{Discussion}
\label{sec:discussion}

\subsection{Number of image systems in a model}
The selection of image systems with or without confirmed redshifts is usually not a choice in building a lens model because, for most clusters, the number of constraints is small regardless and thus modelers require a minimum number of constraints to build a statistically meaningful model. However, the paradigm has changed with the onset of the HFF, where there is a seemingly infinite number of multiple image systems and several spectroscopic redshifts from which to build our models. Statistical errors in these scenarios are now much lower than the systematics, so including or rejecting a candidate image systems in a model is now a question of its influence on systematic error.  Our results show that models reach a threshold in systematic errors across all diagnostics once $N\geq25$ and $n>0$ have been established. For the test models, the identification of images and redshift measurements are known with certainty; however, that is not the case in real scenarios. Beyond this threshold, rejecting an incorrect image system or redshift based on a high degree of uncertainty will likely deflate rather than inflate systematic errors.

\subsection{Finding new multiple image systems}
From \autoref{fig:rms}, we learned that spectroscopic redshift systems are needed to improve the image plane rms of a model when few constraints are available ($N<15$). While this may have its applications for post-modeling, image predictability is most applicable for improving an existing model by using its deflection to find new multiple image systems. The results of this work emphasize the importance of including more spectroscopic redshifts early on in the stages of lens modeling because models built using fewer spectroscopic redshifts have more error in predictability and thus are more likely to find false image systems. While the brightest and largest multiple image systems are obvious by morphology and color without confirmed redshifts, fainter and smaller systems are more ambiguous, especially where many faint galaxies at all redshifts pass the detection limit and could be confused for lensed galaxy candidates.

\subsection{Constraining mass}
Mass profiles of galaxy clusters are quite robust to redshift confirmation of multiple image systems. \autoref{fig:mass} shows that including more image systems in a model with at least a handful of spectroscopic redshifts helps to reduce systematic errors in mass profile. The systematic error on total projected mass out to 1 Mpc is only 2\% for models with $N\geq25$ and $n>0$ (4\% for $N<25$). This result is promising for using strong lensing clusters for cosmology -- future large area surveys will find hundreds of clusters and complete spectroscopic follow-up will not be a feasible task. Knowing that mass within the Einstein radius has low systematic errors will add further significance to cosmological models constrained by strong lensing masses. However, these low errors lie on top of statistical errors and systematics due to structure along the line of sight. We note that we only investigated the mass profile of a single, massive cluster in this work. It would be important in future work to test if this result holds for less massive clusters that lens only a handful of images.

\subsection{Improving magnification estimates}
We find that the regions of the lens map with the highest systematic error are those close to the critical curve and/or along portions with significant curvature where the shear is high and there are few multiple images. The lowest error regions are those covered by multiple images, along portions of the critical curve that are straight. We found that models with low $N$ and low $n$ tend to estimate lower magnifications overall. As we saw, the free parameter redshifts solved for in models with many free parameter redshifts tend to be biased high, which results in a lower mass and thus lower magnification, which matches the trends we see in Figures \ref{fig:dlsds}-\ref{fig:histograms}. While mostly qualitative, this information is useful for anyone questioning the accuracy of a magnification value. While it is trivial to estimate the magnification and statistical uncertainties for a single image plane position by blindly computing it for one pixel in a magnification map, one needs to consider the pixel position within the full image plane to begin estimating the systematics.

\subsection{Models without spectroscopic redshifts}
Since the deflection angle depends on source redshift, the mass estimate within the Einstein radius depends on the redshift of the multiple images. If the redshift of the source is unknown, then the mass is degenerate with redshift. Therefore, lens models need at least one spectroscopic redshift to break the degeneracy. We test this theory by running models without spectroscopic redshifts and find that it is indeed the case that models built with even a handful of spectroscopic redshifts outperform all models built without any spectroscopic redshifts across all of our diagnostics. The models tend to predict redshifts that are higher than truth for nearly every image system, and therefore under-predict the mass by up to 10\% at the Einstein radius, and produce magnifications that can be either highly under- or over-predicted depending on the selection of constraints. When also factoring in statistical errors, which are high for low $N$ systems, and other systematics like structure along the line of sight, any lensing outputs from models with no spectroscopic redshifts should be treated with caution.

\begin{figure*}
\center
\includegraphics[width=0.8\textwidth]{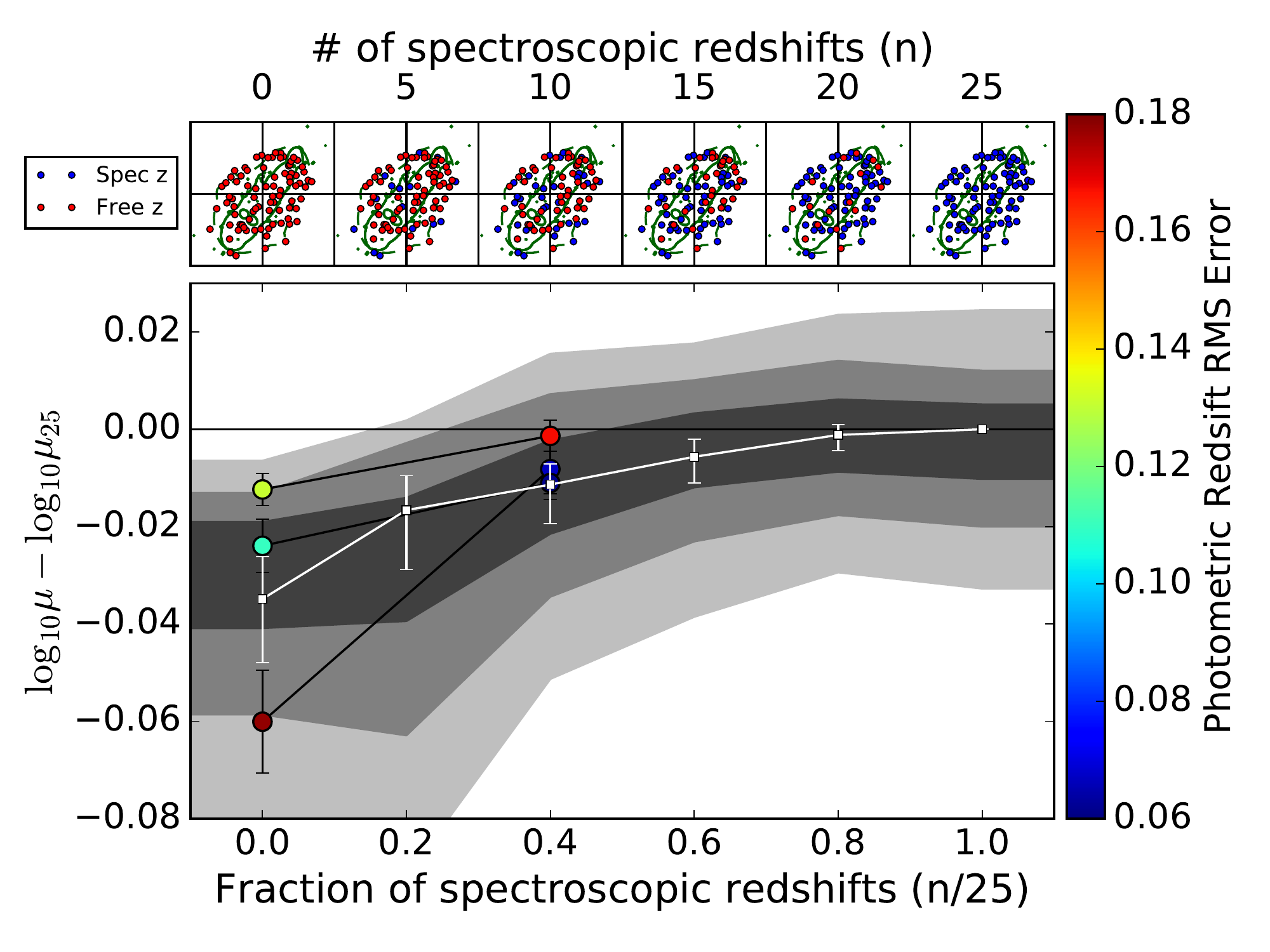}
\caption{Relative magnification error in ($z=2$) for the region of pixels shown in \autoref{fig:hst} vs. the fraction of spectroscopic redshift systems for six lens models built using the same 25 image systems. The first model uses no spectroscopic redshifts, the second model adds spectroscopic redshifts to 5 systems, the third model adds spectroscopic redshifts to 5 more systems (10 total), etc., until all systems have spectroscopic redshifts. The errors are with respect to the model with all 25 spectroscopic redshift systems. The values are the median and $1\sigma$ range of values with the region of the best-fit model. The gray contours represent the 1,2, and 3$\sigma$ statistical errors estimated from the MCMCs. The top row shows the image plane positions of the images from the 25 systems used as constraints. The blue and red points represent systems with and without spectroscopic redshifts, respectively. Each map is 200\arcsec$\times$200\arcsec\ centered on the origin defined in \autoref{fig:hst}. The green solid and dashed lines indicate the locations of the $z=2$ critical curve and the region of multiple images, respectively. The colored circles represent models using the same constraints as the models shown in the maps above; however, photometric redshift measurements are used as the priors for the free parameter redshifts rather than a uniform random prior. The colors indicate the rms error in the photometric redshifts used for that particular model: $(z_\mathrm{spec}-z_\mathrm{phot})/(1+z_\mathrm{spec})$ (see \S~\ref{subsec:photoz}).}
\label{fig:single_compare}
\end{figure*}

\subsection{Increasing the number of spectroscopic redshifts in a single model}
\label{subsec:n_specz}
Our results in \S\ref{subsec:magnification} showed that there was no trend in systematic errors on magnification with the fraction of spectroscopic redshifts used in a model when considering random selections of $n,m$. However, in cluster lensing scenarios similar to the HFF, the selection of image systems mostly stays the same and the fraction of spectroscopic redshifts $n/N$ increases over time as more spectroscopic data are collected. Ongoing lensing analyses of the HFF have so far indicated that increasing the fraction of spectroscopic redshifts for a given cluster may decrease systematic errors on its lens models. The tensions between observations of Supernova Tomas in Abell 2744 and Supernova Refsdal in MACS J1149.6+2223 and the predictions from several different lens models (i.e., magnification, time delays) have weak negative correlations with the fraction of spectroscopic redshifts \citep{Rodney:2015uq,Rodney:2016sf}. In this scenario, the lens models are built by different teams using nearly the exact same identifications of multiple image systems, with some models including new spectroscopic systems in addition to existing sets.

To test whether we see this trend in the simulations, we design a set of lens models that represent a progression in increasing spectroscopic redshift fraction. We construct six new lens models of Ares, each using the same set of 25 image systems. The first model is constructed without any spectroscopic redshifts, the second adds spectroscopic redshifts to 5 of these systems, the third adds an additional five spectroscopic redshifts to the existing 5 (10 total), etc., until all image systems have spectroscopic redshifts. The 25 systems and each additional spectroscopic system are selected carefully in order to maintain a roughly uniform distribution of locations in the image plane and of redshift. \autoref{fig:single_compare} shows the magnification error of these models with respect to the model with $n=N$, in the same manner as \autoref{fig:specz_fraction}. Here, it is clear that the accuracy of the magnification estimates improves with increasing $n/N$, indicating that measuring the spectroscopic redshifts of known lensed galaxies that are used as constraints will help decrease systematic errors while the precision is set based on the total number of systems. This result is consistent with those of \citet{Rodney:2015uq,Rodney:2016sf}, but shows a much stronger correlation. It is likely that the trends in the HFF models are weakened by systematics in the modeling methods themselves and that the selection of constraints were not exactly identical between models.

It is still important to note, however, that the model with $n=N$ is offset in magnification error from the fiducial model by about -0.01 mag. This result is expected, as we saw in \autoref{fig:histograms} that models with $n=25,m=0$ have a systematic error of 0.02 mag with respect to the fiducial. With this in mind, increasing $n/N$ for a single model is most effective at decreasing systematic errors in magnification up to about $n/N\sim0.5$. Beyond that, the exact selection of all the image systems used in the model is a more significant source of systematic error.

While an investigation of the effect of photometric redshift information is beyond the scope of this paper, we do include a test case where we constrain the free redshift parameters with priors from photometric redshift catalogs. This preliminary analysis indicates that photometric redshifts may increase the accuracy of the lens model, but can also result in significantly inaccurate results if not handled with care. We discuss this in \S~\ref{subsec:photoz} below.

\section{Future work}
\label{sec:future}
While we have begun to thoroughly investigate the systematics of lens modeling in this paper, there are still many contributing factors we have not yet explored. Here, we considered how using different random subsets of spectroscopic and free parameter redshift image systems in a strong lens model affects the resulting multiple image predictability, mass profiles, and magnifications. As stated above, these results suggest that the exact selection of constraints and redshift information may be more influential on systematic errors than quantity, especially for the values of the magnification. Thus, we plan to follow-up investigations of constraint selection in our continuation of this work.

\subsection{Observational limits on constraint selection}
We know the selection of multiple image systems is not random by any means and is a function of image brightness, which depends on the source's intrinsic brightness, luminosity distance from observer, and magnification induced by the galaxy cluster. The faintest observed images are less likely to be identified as multiple images. Similarly, obtaining spectroscopic redshifts can depend on image brightness, redshift, and image plane position. Multi-object spectrographs are limited in slit-packing capabilities and may only target the brightest systems for redshift measurements. Spectra of images close to cluster member galaxies might be contaminated, for which may make determining a redshift more difficult. \hst\ grism spectroscopy and integral field spectrographs are able to target many more images; however, they tend to have a limited total field of view. Additionally, completeness of spectroscopic redshifts depends on redshift as bright emission lines get shifted out of the instrument's wavelength coverage for certain redshifts, the so-called ``redshift desert" where redshifts become more challenging to measure. Factoring these selection effects could highlight position and redshift dependencies on the systematic error induced by spectroscopic selection effects.

\subsection{Photometric redshifts}
\label{subsec:photoz}
The current analysis clearly leaves out possible useful information in the form of photometric redshifts; these are typically available for clusters with extensive multiwavelength imaging data. Photometric redshift measurements are prone to their own systematic errors, and while these measurements can become more precise with increased number of bandpasses and deeper data, catastrophic failures can still occur. Photometric redshift measurements can be implemented in the lens modeling process by using the posterior probability distribution for the photometric redshift as the prior for the free parameter redshift in the lens modeling. While we leave a thorough investigation of the affect of photometric redshifts on lensing systematics for future work, we present here a case study.

We re-run the models we used in \S~\ref{subsec:n_specz} with $n=0$ and $n=10$ three more times using different realistic photometric redshift estimates for the priors of images without spectroscopic redshifts. In this experiment, the lensed galaxies used as constraints with spectroscopic redshifts are treated the same as before. However, lensed galaxies without spectroscopic redshifts are not assigned a uniform random prior on their free parameter redshift, but rather a gaussian prior centered on an assumed photometric redshift. We use the ASTRODEEP photometric redshift catalogs for HFF clusters Abell 2744 and MACS J0416 \citep{Castellano:2016lr} to determine our realistic photometric redshift measurements, and supplement the spectroscopic redshift sample of MACS 0416 with the MUSE redshifts measured by \citet{Caminha:2016fk}. We use the spectroscopic and photometric catalogs from this sample to estimate the accuracy of a photometric redshift given its true redshift. For each galaxy in our models of Ares, where we do not include the spectroscopic redshift as a constraint, we draw a galaxy from the ASTRODEEP catalog with a similar spectroscopic redshift to its true redshift in the Ares simulation (within $0.04(1+z_\mathrm{true})$). We then assign the photometric redshift estimate obtained for that ASTRODEEP galaxy as the center of gaussian prior for the Ares galaxy. We assign a typical statistical error on photometric redshifts from the ASTRODEEP catalog as the width of the gaussian prior (these errors can vary significantly from galaxy to galaxy and across redshifts, from a few percent to up to 50\%.) This procedure results in a realistic representation of scatter of photometric redshift values at a fixed spectroscopic redshift, as well as the rate of catastrophic failures in fields that are by construction similar to our simulation. We include these models in \autoref{fig:single_compare} as colored circles, where the color matches the rms error in photometric redshift. The results show that photometric redshifts can improve the lens modeling process, however, only when the photometric redshifts are relatively accurate. As shown in \autoref{fig:single_compare}, using only photometric redshift priors can actually increase systematic errors in magnification over the use of broad uniform random priors when there are catastrophic failures in the photometric redshift measurements. One of the models we used had no spectroscopic redshifts and a photometric redshift rms error of 0.18, including a catastrophic failure with $z_\mathrm{phot} = 1.06$ for $z_\mathrm{spec} = 5.34$. This model produced a worse systematic offset in magnifications compared to the fiducial model. After adding 10 spectroscopic redshifts, the rms error dropped to 0.06, which is likely the result of replacing a catastrophic failure photometric redshift measurement with the true redshift of that system. The resulting model performs slightly better than the model without any photometric redshift information. The two other models with no spectroscopic redshifts had a moderate rms error (0.12) and reduced the systematic error by 0.01-0.02 dex, but did not perform significantly better after adding more spectroscopic redshifts. Photometric redshifts have the highest impact on modeling when there are few to no spectroscopic redshifts; however, this only improves the model if those photometric redshifts are reasonably accurate.

It is worth noting that there are a few aspects of including photometric redshifts in lens modeling that are difficult to simulate because they are highly dependent on the experience of the lens modeler. In the case of the catastrophic failure like in one of the models we tested, it is quite likely that a lens modeler would have rejected any low-redshift solutions that are inconsistent with the lensing geometry based on a model produced by the other images with reasonable photometric redshifts (the $z=5$ critical curve has a much larger extent than the $z=1$ critical curve, so the multiple images should be closer together). We also are not considering that the individual images may have different photometric redshift estimates. Brighter images may have more robust redshifts while fainter or contaminated images may produce wildly different photometric redshifts. A system of images with significantly different photometric redshifts may be less likely to be identified as a system and therefore not used in the lens model. It is unclear at this point how much more photometric redshifts will improve lens modeling if they are not reasonably accurate. However, future purely photometric surveys will have few if no spectroscopic redshifts for the many thousands of lensing clusters predicted to be found. Thus, it is important to investigate how photometric redshifts impact lens modeling and we plan to do so more extensively in future work.

\subsection{Image multiplicity and misidentified multiple images}
Our analysis in this paper investigated how the number of multiple image systems affects the systematic errors in lens modeling. However, we assume that every image system is equal in constraining power and that each system has been correctly identified. In reality, image systems with higher multiplicity (e.g., 4-image systems versus 2-image systems) have a higher weight in the lens modeling. Additionally, higher multiplicity image systems are likely to include radial arcs that will have higher constraining power on the inner slope of the mass profile. From the suite of 350 models we ran, the average image multiplicity (i.e., average number of images per system) ranged from 2.8 to 3.8 for all combinations of $n,m$. We found no trend in systematic errors in the inner and outer slopes of the mass profile nor the magnifications with average image multiplicity. It is possible that trends could occur when the number of image systems is fewer than five, which was the lowest number of systems we investigated. Thus, image multiplicity should be investigated in future work with clusters that have very few multiple image systems to constrain lens models.

We did not account for image identification error in our analysis. The faintest images of a single system are the most likely to be misidentifed because they can be easily confused with other background sources or blended with foreground objects. As stated in \S~\ref{subsec:fiducial_model}, we did not include some of these images that are likely to be misidentified in our models. Therefore, our results show the best-case scenario when lens modelers use only the highest confidence images in their lens models. Simulating the effects of misidentification could be done in future work by comparing models where the faintest image system is perturbed by several arcseconds or not included in the model.

\subsection{Image plane and redshift distribution of lensing constraints}
For models with smaller numbers of image systems, it is important to assess how the spatial distribution of image systems in the image plane affects systematic errors. We found in this work that many of the outlier test models in mass and magnification had uneven spatial distribution of spectroscopic systems. We also saw that models with smaller $N$ had larger spread in mass and magnification because the spatial distribution of the constraints can vary significantly from model to model. As more constraints are added, constraints will more evenly populate the multiple image region. Ares simulates a very massive cluster and it is likely that a cluster of this size will lens more than a handful of image systems. Therefore, we did not attempt to model Ares using fewer than five image systems. We would consider instead modeling a less massive cluster to assess how image plane distribution affects the outcome of a lens model.

As we found in this analysis, the elongated mass distribution of Ares across the sky creates an elliptical critical curve, where the magnification errors are lowest along the straight portions of the critical curve. These elongated mass distributions are common amongst the HFF clusters, which lie at the cusp of mass assembly in the nodes of the cosmic web. It would be interesting to investigate systematics on clusters with more spherical mass distributions. A prime example would be Abell 1689, which has a large number of identified image systems with spectroscopic redshifts and for which existing lens models suggest it has a more circular critical curve \citep{Broadhurst:2005qy,Limousin:2007fk,Coe:2010fy,Diego:2015tg}.

The redshift distribution of lensed sources could potentially increase systematic error as well. The sources used in Ares were well distributed across redshifts from $z\sim0.9-6$; however, this is not the case in reality because the luminosity function of galaxies and the area of the caustic region both depend on redshift. There is an observational bias toward selecting the brightest sources, and as lensing conserves surface brightness, this leads to a higher likelihood of low-redshift sources being identified and used as constraints in a model. We found that some models that were outliers in our analysis for a given $n,m$ had uneven distributions in redshifts. It is well known that multiple redshifts of sources are needed to establish the slope of the mass distribution in a lens model \citep[i.e., break the mass-sheet degeneracy, see][]{Schneider:1995vn}. In cases where all the lensed sources are low redshift, it is possible to extrapolate the mass to larger Einstein radii and thus predict the magnification of higher redshift sources; however, the accuracy of doing so is unknown and is worth investigating in the future.

For clusters such as Ares, built to resemble massive lensing clusters such as the HFF, it is safe to say that there will be a wealth of constraints across the image plane and redshift space. More massive clusters have a larger lensing cross-section and thus have access to a much larger volume of background sources from which to lens. The investigations into image plane and redshift distribution of lensing constraints is best left to more average mass clusters, which will likely only lens a handful of sources. As Ares is too massive to investigate the parameter space of $N<5$, these questions call for a different design in cluster lensing simulations and are best left for future studies.

%==================================================================================
%  CONCLUSION
%==================================================================================
\section{Conclusion}

We have investigated the systematic errors of parametric strong lensing modeling induced by the selection of constraints using our ``unblinded" model of the simulated cluster Ares \citep{Meneghetti:2016xe}. Here we summarize our findings.

\begin{enumerate}
\item The image plane rms based on the full lensing evidence, i.e., the image predictive power of a lens model, improves mostly effectively with increasing number of spectroscopic redshift image systems. This result indicates the necessity for obtaining spectroscopic redshifts early on in the modeling process because they are crucial to increasing the accuracy of finding new multiple image systems. We also have shown, however, that the image plane rms values quoted in the literature, which are computed only from image systems used in the model using best-fit model redshifts, shows the opposite trend. While lower values of the rms computed this way reveal a better model fit with more free parameters and fewer constraints, it can be misleading as a measure of model accuracy.
\item Lens models with at least a handful of spectroscopic redshifts are able to predict the redshifts of image systems without spectroscopic redshifts within 2\% (in $\dls/\ds$); however, they are generally biased higher for image systems with $z>2$.
\item The mass profiles are accurately measured for all variations of lens model constraints with $n>0$: $<4\%$ error within 1 Mpc and $<2\%$ at the cluster Einstein radius.
\item Qualitatively, the magnification error is lowest in regions of the image plane where the multiple images are located and typically along the straight portions of the critical curve. The magnification error is larger at the curved portions of the critical curve and typically biased toward lower values.
\item The accuracy of magnifications increases with total number of image systems, and for $N>20$ plateaus to 2\%. We observe no trend in magnification accuracy with a fraction of spectroscopic redshift when comparing models in which $n,m$ are chosen randomly, as long as this fraction is greater than zero. However, we do find that for a model with a fixed set of multiple images, increasing the fraction of systems with spectroscopic redshifts helps improve the accuracy while maintaining a nearly a constant level of precision.
\item Lens models need at least a few spectroscopic systems in order to produce reasonable estimates of the mass and magnification. Models computed without spectroscopic redshifts are biased toward lower masses ($5-10\%$) and lower magnifications ($>2\%$). The systematic error will be lower for models that use more image systems; however, models built using many image systems without spectroscopic redshifts still produce higher errors than models with only a handful of spectroscopic redshifts.
\item Photometric redshifts can be implemented in lens modeling to improve upon the systematic errors on magnification, especially when there are no spectroscopic data available for the constraints. However, inaccurate photometric redshifts (i.e., catastrophic failures) can actually inflate the systematic errors of a lens model.
\end{enumerate}

Based on our findings, we put forth the following recommendations with regards to strong lens modeling.

\begin{enumerate}
\item  After obtaining new spectroscopic redshifts, newer iterations on existing lens models should be reconstructed using only spectroscopic systems first before including systems with no spectroscopic redshifts as constraints. This method will likely lead to a higher success rate of finding correct multiple image systems.
\item Models with many unknown redshifts report the image plane rms computed for only systems with spectroscopic redshifts as a means to measure model prediction locations of images versus the truth.
\item In circumstances where there are ample numbers of image systems ($N>25$) it may be advantageous to include only the highest confidence image systems (those that are spectroscopically confirmed and/or wholly unambiguous by color and morphology), as removing less confident images may decrease systematics more than the cost of increasing statistical errors.
\item Simply selecting a value from a best-fit model and quoting only statistical errors is not enough for properly estimating magnifications of background sources, one needs to assess the location of that object within the image plane (assuming the source redshift is known) as well and determine whether or not a pure strong lensing analysis is enough to estimate the magnification.
\end{enumerate}

While we have discussed the impact of constraint selection on systematic errors, there are many sources of error we leave to discuss in future papers, i.e., image plane distribution of constraints, redshift, and brightness-dependent selections, photometric redshifts as constraints, unmodeled line-of-sight substructure \citep[e.g.,][]{DAloisio:2014fj}, cluster substructure \citep[e.g.,][]{Limousin:2007fk}, cosmological parameter uncertainty \citep[e.g.,][]{Bayliss:2015lr}, and choice of lensing algorithm \citep{Meneghetti:2016xe}, which we have not quantified here. Additionally, we would like to extend these studies to real clusters in the field and to less massive clusters that represent a larger fraction of the cluster population.

This work is timely as projects such as the lens models for the HFF are continuously being upgraded with several spectroscopic campaigns of multiple image systems underway \citep[ex.,][]{Karman:2015qv,Grillo:2016jk,Jauzac:2016dn,Treu:2016lr}. New and exciting highly magnified cluster-lensed galaxies are being discovered frequently and the demand for precise and accurate lens models has never been higher to understand the lensed background Universe. We hope these results and those to come will help guide and motivate lens modelers and observers alike in meeting those demands and utilize the full power of cosmic telescopes.

\acknowledgements
We thank the anonymous referee for contributing helpful comments and feedback. Many thanks to Massimo Meneghetti, Priyamvada Natarajan, and Dan Coe for facilitating the strong lensing comparison challenge and providing the Ares simulation. We would also like to thank Michael Gladders for suggesting that we explore the parameter space of zero spectroscopic redshifts. This research was supported in part through computational resources and services provided by Advanced Research Computing at the University of Michigan, Ann Arbor.

%\bibliographystyle{../apj}
%\bibliography{../bibdesk_library/lensing_papers}

\end{document}